\documentclass[reprint, prd, amsmath,amssymb, aps,onecolumn,nofootinbib]{revtex4}
\usepackage{graphicx}
\usepackage{hyperref}
\usepackage{dcolumn}
\usepackage{adjustbox}
\usepackage{bm}
\usepackage{aas_macros}
\usepackage{natbib}
\usepackage{xcolor}
\usepackage[thinc]{esdiff}
\usepackage{amsmath}

\begin{document}
\title{Inverse-Compton Scattering of the Cosmic Infrared Background}
\author{Alina Sabyr$^{1}$}
\email{as6131@columbia.edu}
\author{J.~Colin Hill$^{2,3}$}
\author{Boris Bolliet$^{2}$}
\affiliation{$^{1}$Department of Astronomy, Columbia University \\ New York, New York, USA 10027}
\affiliation{$^{2}$Department of Physics, Columbia University \\ New York, NY, USA 10027}
\affiliation{$^{3}$Center for Computational Astrophysics, Flatiron Institute \\ New York, New York, USA 10010}
\date{\today}

\begin{abstract}
The thermal Sunyaev-Zel'dovich (tSZ) effect is the distortion generated in the cosmic microwave background (CMB) spectrum by the inverse-Compton scattering of CMB photons off free, energetic electrons, primarily located in the intracluster medium (ICM).  Cosmic infrared background (CIB) photons from thermal dust emission in star-forming galaxies are expected to undergo the same process.  In this work, we perform the first calculation of the resulting tSZ-like distortion in the CIB. Focusing on the CIB monopole, we use a halo model approach to calculate both the CIB signal and the Compton-$y$ field that generates the distortion.  We self-consistently account for the redshift co-evolution of the CIB and Compton-$y$ fields: they are (partially) sourced by the same dark matter halos, which introduces new aspects to the calculation as compared to the CMB case.  We find that the inverse-Compton distortion to the CIB monopole spectrum has a positive (negative) peak amplitude of $\approx 4$~Jy/sr ($\approx -5$~Jy/sr) at 2260 GHz (940 GHz).  In contrast to the usual tSZ effect, the distortion to the CIB spectrum has two null frequencies, at approximately 196 GHz and 1490 GHz. We perform a Fisher matrix calculation to forecast the detectability of this new distortion signal by future experiments. {\it PIXIE} would have sufficient instrumental sensitivity to detect the signal at $4\sigma$, but foreground contamination reduces the projected signal-to-noise by a factor of $\approx 70$.  A future ESA Voyage 2050 spectrometer could detect the CIB distortion at $\approx 5\sigma$ significance, even after marginalizing over foregrounds.  A measurement of this signal would provide new information on the star formation history of the Universe, and the distortion anisotropies may be accessible by near-future ground-based experiments. 

\end{abstract}
\maketitle
\section{Introduction} \label{sec:intro}
As cosmic microwave background (CMB) photons travel through the universe from the surface of last scattering, they undergo inverse-Compton scattering off of hot, free electrons, located primarily in the intracluster medium (ICM) of galaxy groups and clusters.  This physical process in the late-time universe causes a distortion in the energy spectrum of the CMB, resulting in an increase (decrease) in the higher (lower) frequency range of the spectrum, and is known as the thermal Sunyaev-Zel'dovich (tSZ) effect \cite{SZ1969, SZ1970}.  The spectral dependence of the tSZ effect has a null at $\approx 218$ GHz, where the CMB blackbody photon occupation number is unchanged by the scattering process (the location of this null is altered slightly by relativistic corrections when the scattering electrons are at high temperatures~\cite{Sazonov1998,Challinor1999,Nozawa2006,Chluba2005, Chluba2012}).  The tSZ effect has now become a routine probe with which to detect galaxy clusters (e.g.,~\cite{PlanckClusters2016,Bleem2015,Hilton2021}) and study the thermal electron pressure distribution in the late-time universe using component-separated tSZ maps (e.g.,~\cite{Planck2016ymap,MHN2020,Bleem2021}).  For reviews of the tSZ effect and some of its cosmological and astrophysical applications, see \cite{Rephaeli1995,Birkinshaw1999,ks2002,Carlstrom2002,hp2013,b2018,Mroczkowski2019}.  

While the tSZ effect refers to the spectral distortion in the CMB spectrum, it is generally the case that any photon distribution will undergo Comptonization upon scattering with free electrons~\cite{Blumenthal1970,Rybicki1986}, e.g., as seen in X-ray sources~(e.g.,~\cite{Pozdnyakov1983,Payne1980}), synchrotron emission from active galactic nuclei~(e.g.,~\cite{Maraschi1992,Konopelko2003}), or in the lobes of radio galaxies~(e.g.,~\cite{Harris1994,Tashiro2001,Brunetti2001}).  In the context of cosmology, inverse-Compton scattering of the CMB (i.e., the tSZ effect) has received the most study, but other photon backgrounds are also expected to undergo the same scattering process, including the cosmic radio background~\cite{HolderChluba2021,Lee2021}, cosmic 21 cm radiation \cite{Cooray2006}, and the cosmic infrared background (CIB), the latter of which we focus on here.  Like the CMB, the photons in the radio and infrared backgrounds are at much lower energies than typical electrons in the ICM ($T_{\gamma} \ll T_e \sim$ keV), and thus Comptonization leads to the photons being upscattered to higher energies.  However, the exact shape of the inverse-Compton-induced distortion will differ in each case, as the incoming photon spectral energy distribution (SED) is different for the CMB, CIB, or radio photons.  For the CMB, the blackbody nature of the incoming photons significantly simplifies the calculation \cite{SZ1969, SZ1970}, while results for a power-law radio SED have recently been presented in \cite{HolderChluba2021} and \cite{Lee2021}. In this work, we consider the case of a realistic dusty galaxy SED for the incoming photons, modeled as a modified blackbody at low frequencies combined with a power-law decay at high frequencies, as appropriate for the star-forming galaxies that generate the CIB. We also treat in detail the co-evolution of the Compton-$y$ (scattering electrons) and CIB (incoming photons) fields. Although the CIB distortion has been qualitatively discussed \cite{HolderChluba2021, Lee2021} and estimated previously using a different CIB model \cite{Kholupenko2015}, no calculation using a halo model approach was performed.

The CIB is a diffuse radiation field originating from dust in star-forming galaxies, which re-emit the absorbed ultraviolet light from stars in the infrared \cite{Bond1986,Dwek1998}. For this reason, the CIB is a valuable probe of both large-scale structure and the star formation history of the Universe \cite{Song2003,Holder2013,Planck2014lensing,Kennicutt1998,Knox2001}. It was first observed by the Cosmic Background Explorer (COBE) mission \cite{Puget1996, Dwek1998, Fixsen1998} and has since been measured at high precision by many experiments including IRIS \cite{MivilleDesch2005, Penin2012}, \emph{Herschel} \cite{Amblard2011, Viero2013}, \emph{Planck} \cite{Planck2014}, ACT \cite{Choi2020}, and SPT \cite{Hall2010}.

One important distinction between the CIB and the CMB is that the latter is truly a background photon field, in the sense that it was produced at very high redshift (during or shortly after post-inflationary reheating); thus, new CMB photons are not produced at the redshifts where galaxy groups and clusters are forming ($z \lesssim 3$).  Therefore, the scattering electrons in the ICM of each cluster all see the same monopole CMB blackbody SED, with the temperature simply scaling as $\propto (1+z)$.  The scattered photons cool adiabatically in the same fashion as the unscattered photons, which leads to the famous redshift independence of the tSZ effect: the tSZ spectral dependence is the same for a cluster at $z=1$ as for a cluster at $z=0.001$.\footnote{The only redshift dependence of the tSZ effect comes from the changes in the properties of the clusters at different redshifts (e.g., typical mass and size). As a result, for example, the catalogues of clusters detected through the tSZ signal are not redshift-independent, particularly for experiments with relatively large beams (e.g. \cite{PlanckClusters2016})}  In contrast, the photons that comprise the CIB are produced at cosmologically late times ($z \lesssim 6$; see Fig.~\ref{fig:icib_didz}), which overlap (partially) with the epoch at which groups and clusters are forming.  Thus, the monopole CIB SED seen by ICM electrons in a cluster at $z=1$ is different than that seen by ICM electrons in a cluster at $z=0.001$, which renders the calculation of the inverse-Compton effect on the CIB somewhat more complicated than that for the CMB.\footnote{Previous work on the CIB included the redshift dependence of the CIB but not of the Compton-$y$ field \cite{Kholupenko2015}. The same issue will be present for inverse-Compton scattering of the cosmic radio background. Approximating the cosmic radio background as a ``true background'' may be more valid than for the CIB, as the origin of the radio background is much less well-understood~\cite{Seiffert2011,Singal2018}. Nevertheless, some fraction of the radio background must be generated by radio sources at late times, which will overlap in redshift with the groups and clusters that comprise the Compton-$y$ field. Previous works on the distortion in the radio background emphasized the importance of this redshift-dependence, and although detailed modeling was neglected, included an overall $f(z)$ factor to illustrate some of the possible effects of the redshift dependence ~\cite{HolderChluba2021,Lee2021}.}

In this work, we treat this effect in detail (for the first time, to our knowledge) by making use of a complete halo model formalism for both the CIB and Compton-$y$ fields.  We focus solely on the monopole sky signal, leaving the calculation of anisotropies to future work (apart from a brief estimate of the power spectrum in Appendix~\ref{app:as}).  We compute the CIB monopole using a standard halo model prescription, in which infrared galaxies with a specified SED (a modified blackbody at low frequencies combined with a declining power-law at high frequencies) are assigned to dark matter halos as a function of halo mass and redshift.  The Compton-$y$ field is constructed simultaneously using a prescription for the thermal electron pressure profile as a function of halo mass and redshift.  We then compute the CIB monopole that is scattered by the ICM electrons in halos at each redshift, and use this in the Kompaneets equation to compute the differential contribution to the total inverse-Compton distortion.  Taking the integral over all redshifts, we obtain the total inverse-Compton distortion to the observed CIB monopole at $z=0$, accounting self-consistently for the co-evolution of the CIB and Compton-$y$ fields.  As is already well-known, the CIB monopole at $z=0$ is well-approximated at low- to mid-frequencies by a simple modified blackbody SED, which we use as a toy model to compare to our detailed numerical calculations using the halo model.  The toy model captures the order of magnitude of the distortion signal, but the detailed shape of the spectral distortion is only described correctly by the full halo model calculation.  We consider only the lowest-order distortion to the CIB monopole here (the analogue of the tSZ effect in the CMB), leaving a full treatment of kinematic and relativistic effects (along the lines of e.g.,~\cite{Chluba2012} and \cite{Lee2021}) to future work.

The resulting inverse-Compton CIB distortion represents a new signal that can be detected by monopole experiments (e.g. {\it PIXIE} \cite{Kogut2011}) or by anisotropy experiments (e.g., CCAT-p~\cite{CCATp2021}), although further theoretical calculations are needed to compute the latter observables (we provide a first estimate of the power spectrum in Appendix~\ref{app:as}).  In addition, separating the inverse-Compton CIB distortion anisotropies from other signals in the small-scale far-infrared sky will require dedicated analysis.  A detection of the monopole signal considered here would yield new information about the redshift kernel of the CIB, as well as the SED properties of the CIB sources.  This signal is also potentially a  new ``foreground'' for CMB spectral distortions, such as the all-sky relativistic tSZ distortion~\cite{Hill2015,Abitbol2017,Thiele2022}.  As we show below, the impact on forecasts for CMB spectral distortions' signal-to-noise is fortunately small.

The remainder of this paper is organized as follows.  In Sec.~\ref{sec:theory}, we present the theory underlying our calculations, including the inverse-Compton scattering of a toy-model modified blackbody SED, the halo model formalism for the CIB and Compton-$y$ fields, and the full, self-consistent calculation of the CIB monopole and its inverse-Compton-induced distortion.  In Sec.~\ref{sec:results}, we present detailed numerical results for the expected monopole distortion signal, including its dependence on physical CIB parameters, and determine its detectability in near-future experiments and its impact on CMB spectral distortion forecasts.  We discuss our results and conclude in Sec.~\ref{sec:conclusion}.

We assume a flat $\Lambda$CDM cosmology with parameters consistent with those from \emph{Planck}~\cite{Planck2018parameters}: $\Omega_m = 0.316$, $\Omega_b = 0.049$, $\sigma_8 = 0.811$, $n_s = 0.966$, and $H_0 = 67.3$ km/s/Mpc. 
\section{Theory} \label{sec:theory}
In this section, we compute the thermal SZ distortion of the sky-averaged CIB, in the non-relativistic approximation.  We consider two different models for the CIB monopole: (i) a simple modified blackbody SED (Sec.~\ref{ss:tm}), which we use as a toy model to gain intuition for the expected distortion signal, and (ii) a complete halo model calculation, including simultaneous treatment of the Compton-$y$ field (Sec.~\ref{sec:classtheory} and~\ref{ss:dicib}).
\subsection{Toy Model: Modified Blackbody SED}\label{ss:tm}
In the toy model calculation, we treat the CIB as a ``true background'' and assume that all the Compton-$y$ sources are concentrated at $z=0$. With these approximations, we are able to compute the distortion generated by all scatterers on the full past lightcone in a single operation. In the full, halo model calculation approach described later in Sec.~\ref{sec:classtheory}, however, we account for the simultaneous redshift evolution of both the CIB and Compton-$y$ fields.

The CIB monopole emission for photon frequencies 100 $\lesssim \nu \lesssim$ 1000 GHz is described relatively well by a modified blackbody (MBB) spectrum \cite{Planck2014}. 
Therefore, for an initial toy model calculation of the inverse-Compton effect in the CIB, we adopt an MBB photon occupation distribution:
\begin{equation}
\label{eqn:modifiedN}
N_{\rm MBB}(\nu) = A_{\rm MBB}\left(\frac{\nu}{\nu_{0}}\right)^{\beta}\frac{1}{e^\frac{h\nu}{k_{\rm B}T_{\rm MBB}}-1}
\end{equation}
with a corresponding specific intensity 
\begin{equation}
\label{eqn:modifiedI}
    I_{\rm MBB}(\nu) = \frac{2h\nu^{3}}{c^{2}}N_{\rm MBB}(\nu) \,,
\end{equation}

\noindent where $N_\mathrm{MBB}(\nu)$ is the photon occupation number at photon frequency $\nu$, $A_{\rm MBB}$ is an overall amplitude parameter, $\beta$ is the emissivity index, $\nu_{0}$ is the normalization frequency, $T_{\rm MBB}$ is an effective temperature, $k_{\rm B}$ is Boltzmann's constant, and $h$ is Planck's constant. We note that $T_{\rm MBB}$, $\beta$, and $\nu_{0}$ are not physical quantities of any photon source, but simply free parameters of this function, generally determined via a fit to data.
In this initial toy model calculation, we approximate the CIB monopole at $z=0$ using Eq.~\ref{eqn:modifiedI} (the numerical values of the free parameters will be discussed below).  Using the Kompaneets equation, we then compute the inverse-Compton-induced distortion that arises from CIB photons obeying this SED scattering off free electrons in a cluster at $z=0$.

Multiplying the resulting distortion template by the total Compton-$y$ value expected in our Hubble volume, $y \approx 1-2 \times 10^{-6}$~\cite{Hill2015,Dolag2016,chiang2020,Thiele2022}, then gives a rough estimate of the observable CIB distortion signal. (Our fiducial model, described in the next subsection, yields $y = 1.58\times10^{-6}$).  The correct, complete calculation of the inverse-Compton CIB distortion, which we perform in the next subsection, computes the CIB monopole SED seen by clusters at each redshift along our past lightcone, and computes the distortion induced by the Compton-$y$ field at that redshift.

In the non-relativistic limit ($k_{\rm B} T_e \ll m_e c^2$), the interaction between photons and electrons can be described by the Kompaneets equation \cite{Kompaneets1957}, which in the limit that $T_{\gamma} \ll T_{e}$ reduces to  
\begin{equation}
\label{eqn:kompaneets}
    \diffp{N}{y}\approx \frac{1}{\nu^{2}}\diffp{}{\nu}\left[\nu^{4}\diffp{N}{\nu}\right] \,.
\end{equation}
Here, $y$ is the dimensionless Compton-$y$ parameter, which determines the strength of the Comptonization, and the derivative is understood to be taken along the photon path, i.e., the equation expresses the change in the photon occupation number as the photons traverse a free electron gas, as a function of the Compton-$y$ distribution along the photon path.  The $y$-parameter is defined as
\begin{equation}
\label{eqn:y}
    y(\hat{\textbf{n}})=\frac{\sigma_{T}}{m_{e}c^{2}}\int dl \, n_{e}(\hat{\textbf{n}},l)k_{B}T_{e}(\hat{\textbf{n}},l)= \frac{\sigma_{T}}{m_{e}c^{2}}\int dl P_e(\hat{\textbf{n}},l)
\end{equation}
where $\sigma_{T}$ is the Thomson cross-section, $m_{e}c^{2}$ is the electron rest-mass energy, $n_{e}$, $T_{e}$, and $P_{e}$ are respectively the electron number density, temperature, and pressure, $\hat{\textbf{n}}$ is the position on the sky, and $l$ is the distance along the line of sight (LOS). (Our computation of $y$ within the halo model is presented in Sec.~\ref{sec:classtheory}.)

Let us first recall the resulting distortion in the case of the CMB, in which the initial photon occupation number follows the blackbody form. Assuming  $y\ll1$ (i.e., the single scattering limit for an optically thin cloud of electrons), using a blackbody photon distribution in the Kompaneets equation yields the well-known thermal SZ distortion \cite{SZ1969}:

\begin{equation}
\label{eqn:CMBdN}
    \Delta N_{\rm CMB}(x_{_{\rm CMB}}) = y \, \mathcal{G}(x_{_{\rm CMB}})\quad \mathrm{with}\quad x_{_{\rm CMB}}\equiv \frac{h\nu}{k_\mathrm{B}T_{\rm CMB}}\quad\mathrm{and} \quad\mathcal{G}(x)\equiv \frac{x e^{x}}{(e^{x}-1)^{2}} \left( x \, {\rm coth} \left( \frac{x}{2} \right) - 4 \right)
\end{equation}
\noindent where the approximation $\partial N /\partial y\approx \Delta N / y$, valid for small Compton-$y$, was used. 

Now turning to the CIB, using Eq.~\ref{eqn:modifiedN} in the Kompaneets equation, we obtain the change in photon occupation number for the toy model MBB CIB SED:

\begin{equation}
\label{eqn:modifieddN}
\Delta N_\mathrm{MBB}(x) = y A_{\rm MBB} \left(\frac{x}{x_{0}}\right)^{\beta}\left[\frac{\beta(\beta+3)}{e^{x}-1}-\frac{ x(2\beta+4+x)e^{x}}{(e^{x}-1)^{2}}+\frac{2x^{2}e^{2x}}{(e^{x}-1)^{3}}\right] \quad \mathrm{with}\quad x\equiv \frac{h\nu}{k_\mathrm{B}T_{\rm MBB}} \,.
\end{equation}

\noindent Note that the relevant temperature in the dimensionless frequency $x$ is now $T_\mathrm{MBB}$ rather than $T_{\rm CMB}$, and we have implicitly assumed that $T_\mathrm{MBB}$ scales like $(1+z)$ in the same way that $T_{\rm CMB}$ does. Eq.~\ref{eqn:modifieddN} can also be rewritten in terms of the blackbody tSZ distortion function $\mathcal{G}(x)$ as

\begin{equation}
    \Delta N_{\rm MBB}(x) = y A_{\rm MBB} \left(\frac{x}{x_{0}}\right)^{\beta}\left[\mathcal{G}(x)+\frac{\beta}{e^{x}-1}\left(\beta+3-\frac{2xe^x}{e^{x}-1}\right)\right] \,.
\end{equation}

When $\beta=0$, we recover the spectral distortion for a blackbody spectrum found in Eq.~\ref{eqn:CMBdN}, as expected.  We also note that the limits of $\nu^{4}N(\nu)$ and $\nu^{4}dN(\nu)/d\nu$ need to vanish as $\nu \rightarrow 0$ and $\nu \rightarrow \infty$ so that the total number density of photons is conserved over time \cite{Weinberg2008}. Requiring that the limit of $\nu^{4}N_{\rm MBB}(\nu)$ vanishes as $\nu \rightarrow 0$ sets a constraint that $\beta > -3 $ and requiring that $\nu^{4}dN_{\rm MBB}(\nu)/d\nu$ vanishes as $\nu \rightarrow 0$ sets the constraint $\beta > -2$.\footnote{In practice, setting constraints on $\beta$ using these limits is more complicated. For example, free-free absorption as $\nu \rightarrow 0$ means that there will be a finite number of photons regardless of what value $\beta$ is.} This mathematical constraint for no production or destruction of photons during scattering is consistent with the positive power-law indices for modified blackbody SEDs due to infrared sources in the literature, e.g., $\beta \approx 1.75$ \cite{Planck2014}.
Converting from photon occupation number to the observable CIB spectrum, our toy model spectral distortion is thus
\begin{equation}
    \label{eq:modified_dI}
    \Delta I_{\rm MBB }(\nu) = \frac{2h\nu^{3}}{c^{2}}\Delta N_{\rm MBB}\left( x \right) \,,
\end{equation}
with $\Delta N_{\rm MBB}(x)$ given by Eq.~\ref{eqn:modifieddN}.

To set the numerical values of the parameters appearing in Eqs.~\ref{eqn:modifiedN} and~\ref{eqn:modifiedI}, we compare the simple MBB SED to the results of our detailed halo model calculation of the CIB monopole (see Sec.~\ref{sec:classtheory}).  From our halo model implementation we find that the CIB monopole peak amplitude is $\approx 1.1 \times 10^{6}$ Jy/sr at frequency $\approx 1290$ 
GHz (see Fig.~\ref{fig:icib_didz}).  We show the best-fit MBB SED to the halo model-computed CIB monopole in Fig.~\ref{fig:matchI}. To perform the fit, we fix $\nu_{0} = 353$ GHz and use the {\tt curve\_fit} function from the SciPy library \cite{2020SciPy} to find the best-fit values of the parameters. Fitting the monopole SED between 100 GHz and 1 THz, we obtain $T_{\rm MBB} = 12.74$ K, $\beta = 1.49$, and $A_{\rm MBB} = 6.1 \times 10^{-6} \, {\rm sr}^{-1}$.
We note that there is a significant degeneracy between $\beta$ and $T_{\rm MBB}$ in the fit.  In addition, as seen in Fig.~\ref{fig:matchI}, the MBB toy model fails to describe the true CIB monopole at $\nu \gg 1$ THz.  Nevertheless, it is still useful to gain intuition for the expected distortion signal, and it is quite accurate at $\nu < 1$ THz.

Fig.~\ref{fig:cibdistortion} (left panel) shows the toy model calculation for the CIB distortion signal, computed using Eq.~\ref{eq:modified_dI} with the best-fit MBB parameters discussed in the previous paragraph.  To obtain the amplitude of the distortion, we use the total $y$ value obtained in our halo model calculation in the next subsection, $y = 1.58 \times 10^{-6}$ (Fig.~\ref{fig:icib_didz}). Thus, making the simplifying assumption that the entire distortion signal is generated at $z=0$, i.e., that the distortion signal is proportional to the total Compton-$y$ multiplied by the change in the MBB photon distribution (Eq.~\ref{eqn:modifieddN}), we obtain the black curve in the left panel of Fig.~\ref{fig:cibdistortion}.  The amplitude of the distortion agrees with a back-of-the-envelope estimate: since the CIB monopole amplitude is $\sim 10^{6}$ Jy/sr and $y \sim 10^{-6}$ we expect that the distortion has an amplitude of $\sim$ few Jy/sr.

We can use the results of the toy model calculation to determine the approximate null frequencies of the inverse-Compton CIB distortion, analogous to the null frequency of the usual tSZ effect at 218 GHz.  The null frequencies occur where Eq.~\ref{eqn:modifieddN} vanishes, which yields:
\begin{equation}
    \label{eq:zerodN}
    \frac{x^{\beta}}{(e^{x}-1)^{3}}\left[\beta(\beta+3)(e^{x}-1)^2-xe^x(2\beta+4+x)(e^x-1)+2e^{2x}x^{2}\right] = 0 \,.
\end{equation}
In the limit that $x\gg1$, such that $e^x-1 \approx e^x$, this yields
\begin{equation}
    \label{eq:highernull}
    x_{\rm null} \approx 2+\beta \pm \sqrt{4+\beta} \,,
\end{equation}
while in the limit of $x\ll1$, writing $e^x = 1+x + \frac{x^2}{2}+O(x^3)$, we find the roots
\begin{equation}
    \label{eq:lowernull}
    x_{\rm null} \approx \frac{3\beta^2+9\beta\pm\sqrt{96+72\beta-3\beta^2-18\beta^3-3\beta^4}}{4+5\beta+\beta^2} \,.
\end{equation}
For our fiducial value of $\beta=1.49$, the null frequencies in these limits are $x_{\rm null} \in \{1.15, 5.83\}$ and $x_{\rm null}\in\{0.66,2.28\}$, respectively.  However, in the first case it is clear that the assumption $x \gg 1$ only holds for the root $x_{\rm null} \approx 5.83$.  Manifestly, the approximate physical root is thus $x_{\rm null} \approx 2+\beta+\sqrt{4+\beta}$. Similarly, in the second case, the assumption $x \ll 1$ holds only for $x_{\rm null} \approx 0.66$, which is the negative branch of the quadratic solution.  We thus anticipate two physical null frequencies, and we expect that the exact numerical values of the null frequencies should be near these approximate results.  Indeed, by directly solving Eq.~\ref{eq:zerodN} numerically, we find the roots $x_{\rm null} \simeq 0.65$  and $x_{\rm null} \simeq 5.79$, which confirms our expectation.  

Eqs.~\ref{eq:highernull} and~\ref{eq:lowernull} demonstrate that the null frequencies are determined by the emissivity index $\beta$ (and of course the temperature $T_{\rm MBB}$ that converts from $x$ to physical frequency).  The special case of $\beta=0$, corresponding to a pure blackbody SED, yields approximate null frequencies in the $x \gg 1$ and $x \ll 1$ limits at $x_{\rm null} \approx 4$ and $x_{\rm null} \approx -2.45$, respectively, with the latter clearly being unphysical.  The former, however, is indeed the classic tSZ null frequency at $\nu \approx 220$ GHz (the detailed numerical solution in this case yields $x_{\rm null} \simeq 3.83$, i.e., $\nu_{\rm null} \simeq 218$ GHz for $T_{\rm CMB} = 2.726$ K). For our fiducial value of $T_{\rm MBB} = 12.74$ K, the roots $x_{\rm null} \simeq 0.65$ and $x_{\rm null} \simeq 5.79$ correspond to $\nu_{\rm null} \simeq$ 174 GHz and 1540 GHz, respectively. 

The null frequencies can be seen in Fig.~\ref{fig:cibdistortion}.  It is interesting to note that unlike in the standard CMB case, we find that there are two null frequencies in the CIB inverse-Compton distortion.  The higher null frequency (1540 GHz) arises near the peak of the CIB monopole, analogous to the usual zero-crossing in the blackbody CMB tSZ effect at $\nu \simeq 218$ GHz.  At this null, lower-frequency photons below the SED peak are upscattered to higher frequencies above the peak. The unique lower null frequency at 174 GHz arises due to the steepness of the CIB monopole SED at low frequencies and the ``tilt'' in the spectrum due to broadening of the SED generated by Compton scattering. In fact, using the results of Eq.~\ref{eq:lowernull}, we can determine a condition on $\beta$ for the existence of this lower null frequency.  Requiring the root in this limit to be physical ($x_{\rm null} > 0$) dictates that we must have $\beta > 1$ in order for this null frequency to exist.  This can also be validated in detail by numerically solving Eq.~\ref{eq:zerodN} for various $\beta$ values.  (Note that we consider only $\beta \geq 0$ throughout this discussion, as this is the case relevant for dust emission and the CMB.)  If the SED is too shallow ($0 \leq \beta \leq 1$), such as in the case of the CMB blackbody, then the upscattering does not generate a sufficient tilt in the distorted SED to obtain a null frequency.  In Sec.~\ref{ss:comp}, we discuss the null frequencies for the detailed halo model calculation of the inverse-Compton CIB distortion, which are close to (but slightly different from) the approximate results for the toy model MBB SED found here. 

\subsection{CIB and Compton-$y$ within the Halo Model}\label{sec:classtheory}
The toy model of the previous subsection uses a single-temperature MBB SED to approximate the CIB monopole at $z=0$.   
However, this is clearly not a fully realistic model, as we have not actually calculated the thermal emission from star-forming galaxies in detail.  Thus, in this subsection we implement more sophisticated models that provide a good fit to observed dusty galaxy SEDs (see \cite{adb2013} and references therein). Here, for our implementation of the CIB in the halo model (see \cite{CooraySheth2002} for a review of the halo model), we use the CIB model introduced in \cite{Shang2012} (reviewed in detail in \cite{McCarthy2021}) and parameters from \emph{Planck} fits \cite{Planck2014}, which are enumerated in Table~\ref{tab:cibparameters}. 
We briefly discuss alternative CIB models in Sec.~\ref{sec:conclusion}.

In our halo model of the CIB, the galaxy SEDs as a function of galaxy rest-frame frequency $\nu$ and redshift $z$ are represented by a MBB at low frequencies and a power law at high frequencies, and are normalized according to $\Theta(\tilde{\nu})=1$, where
\begin{equation}
\label{eq:SED}
   \Theta(\nu,z)= 
   \begin{cases}
   \left(\frac{\nu}{\tilde{\nu}}\right)^{\beta}\frac{B_{\nu}(T_{\rm d}(z))}{B_{\tilde{\nu}}(T_{\rm d}(z))}\quad\mathrm{for}\quad\nu < \tilde{\nu}\\
   \left(\frac{\nu}{\tilde{\nu}}\right)^{-\gamma} \quad\quad\quad\quad \,\,\,\, \mathrm{for}\quad \nu \geq \tilde{\nu} \,,
   \end{cases}
\end{equation}
where $B_{\nu}(T_d) = \frac{2 h \nu^3}{c^2} \left(e^{\frac{h\nu}{k_\mathrm{B}T_d}}-1\right)^{-1}$ is the blackbody SED at temperature $T_{\rm d}(z)$ parametrized as
\begin{equation}
   T_{\rm d}(z) = T_{0}(1+z)^{\alpha} \,,
\end{equation}
with $\alpha=0.36$ and $T_{0}=24.4$ K in our fiducial model. The pivot frequency $\tilde{\nu}$ is defined via the continuity condition $(\mathrm{d}\ln\Theta/\mathrm{d}\ln \nu)|_{\tilde{\nu}} =-\gamma$, yielding
\begin{equation}
    \tilde{\nu}(z) = \frac{k_\mathrm{B}T_{\rm d}(z)}{h}(3+\beta+\gamma+W_0(\lambda))\quad\mathrm{with}\quad\lambda =-(3+\beta+\gamma)e^{-(3+\beta+\gamma)} \,,
\end{equation}
where $W_0$ is the Lambert function, and with $\beta=1.75$ and $\gamma=1.7$ in our fiducial model. 

\begin{table}[t]
\caption{\label{tab:cibparameters} Fiducial CIB halo model parameters, adopted from \cite{McCarthy2021} and~\cite{Planck2014}. The top six parameter values are from fits to \emph{Planck} spectra, while the bottom two are unconstrained, fixed values (also marked with *).}
\begin{ruledtabular}
\begin{tabular}{ccc}
Parameter & Parameter Description & Value\\
\colrule
$\alpha$ & Redshift evolution of dust temperature & 0.36\\
$T_{0}$ & Dust temperature at $z=0$ & 24.4 K\\
$\beta$ & Emissivity index of SED & 1.75\\
$\gamma$ & Power law index of SED at high frequency&1.7\\
$\delta$ & Redshift evolution of $L$\textendash $M$ normalization & 3.6\\
$M_{\rm eff}$&Most efficient halo mass & $10^{12.6} M_{\odot}$\\
$L_{0}$&Normalization of $L$\textendash $M$ relation & $6.4 \times 10^{-8}$ Jy Mpc$^{2}$/$M_{\odot}$\\
${\rm M_{\rm min}^\mathrm{_{HOD}}}^{*}$ & Minimum halo mass to host a galaxy & $10^{10} M_{\odot}$ \\
$\sigma^{2*}_{\rm L/M}$ & L/M dispersion  & 0.5\\
\end{tabular}
\end{ruledtabular}
\end{table}

Given these SEDs, we compute a galaxy luminosity by making the assumption that it depends simply on the mass $M$ and redshift $z$ of its host dark matter halo, via the functional form 
\begin{equation}
\label{eq:Lgal}
    L^\mathrm{gal}_{\nu} (M, z) =L_{0}\Phi(z)\Sigma(M)\Theta(\nu, z) \,,
\end{equation}
where $L_{0}=6.4 \times 10^{-8}$ Jy Mpc$^{2}$/M$_{\odot}$ is a normalization factor~\cite{McCarthy2021}, $\Phi(z)=(1+z)^\delta$  with $\delta = 3.6$  specifies the redshift dependence of the mass-luminosity relation, and $\Sigma(M)$ is a log-normal distribution 
\begin{equation}
\label{eqn:lognormal}
    \Sigma (M) = \frac{M}{\sqrt{2\pi\sigma^{2}_{\rm L/M}}}e^{-(\log_{10}(M/M_\odot)-\log_{10}(M_{\rm eff}/M_\odot))^{2}/2\sigma^{2}_{\rm L/M}}
\end{equation}

\noindent with dispersion $\sigma_\mathrm{L/M}^2=0.5$ and mean $\log_{10}M_\mathrm{eff}/M_\odot=12.61$.  The latter parameters control the mass dependence, i.e., they determine the range of halo masses responsible for the IR emission. This mass-luminosity relation was introduced in \cite{Shang2012} to account for the suppression of star formation at low and high halo masses due to astrophysical processes such as feedback.

\begin{figure}[!tbp]
\includegraphics[trim=110pt 0pt 110pt 0pt, width=\textwidth]{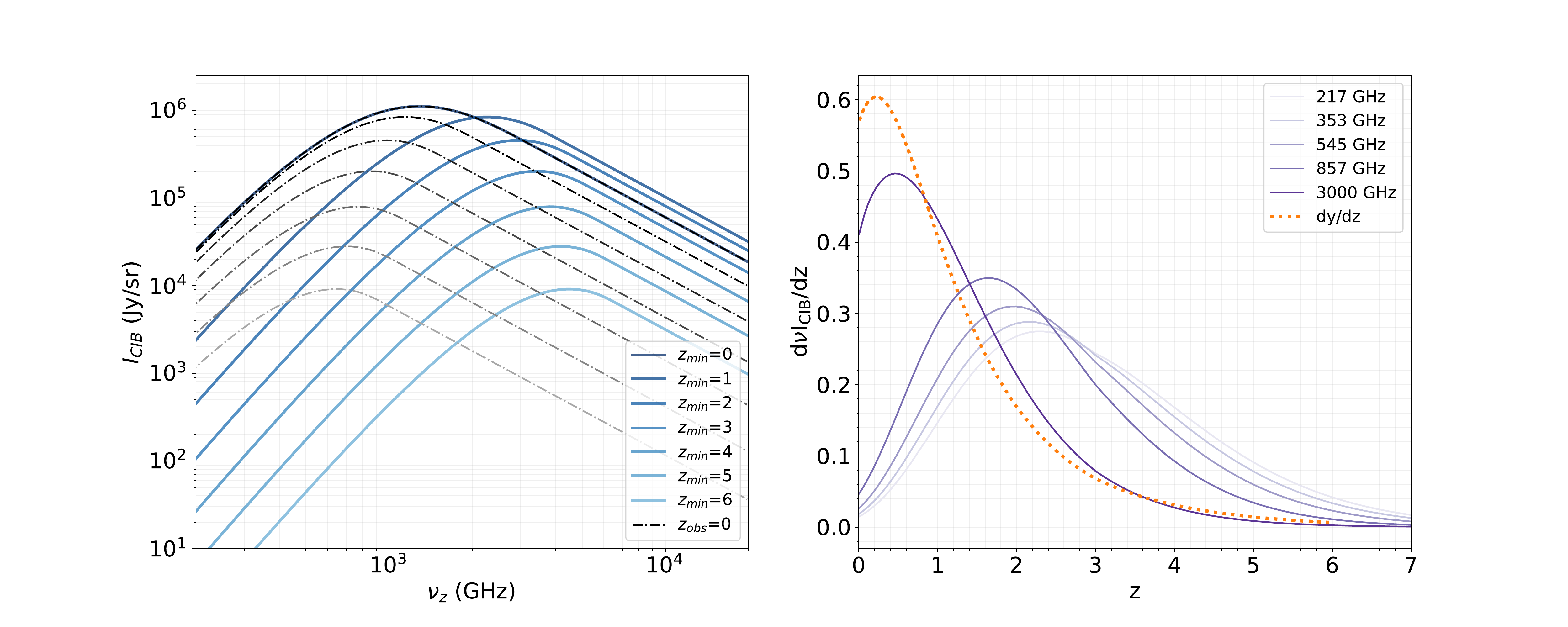}
\caption{\label{fig:icib_didz}\emph{Left:} CIB monopole signal as a function of different minimum source redshifts $z_{\rm min}$, integrated up to $z_{\rm max}=12$ (darker shade corresponding to lower $z_{\rm min}$). The intensity of the full signal at $z=0$ peaks at $I_{\rm CIB} \approx 1.1\times 10^{6}$ Jy/sr near $\nu \approx 1265$ GHz. The grey dash-dotted curves show the CIB monopole as seen by an observer at $z=0$ while the solid blue lines show the signal seen by an observer at $z_{\rm min}$. As expected, the latter are shifted to higher frequencies on the plot, but the two curves coincide for $z_{\rm min}=0$.
\emph{Right:} Normalized CIB redshift kernels at different frequencies (purple, with darker shade corresponding to higher frequency) compared to the redshift kernel of the Compton-$y$ field (dotted). The CIB emission primarily originates between $z\sim1-6$, depending on the frequency under consideration, while the contributions to Compton-$y$ are dominated by $z \lesssim 2$.  The partial overlap of the Compton-$y$ and CIB source kernels implies that their co-evolution needs to be taken into account when calculating the inverse-Compton distortion in the CIB.}
\end{figure}
Assuming that the luminosity of a galaxy is described by the same function ($L^{\rm gal}_{\nu}$ of Eq.~\ref{eq:Lgal}) for both central and satellite galaxies, the central galaxy luminosity can be written as 
\begin{equation}
    L^{\rm cen}_{\nu}(M,z)=N^{\rm cen}(M,z)L^{\rm gal}_{\nu}(M,z)
\end{equation}
where $N^{\rm cen}$ is the expectation value for the number of central galaxies residing in a halo of mass $M$, represented by a step function 
\begin{equation}
    N^{\rm cen}(M, z) = 
    \begin{cases}
    0 \quad \mathrm{for} \quad  M < M_{\rm min}^{_\mathrm{HOD}}\\
    1 \quad \mathrm{for} \quad   M \geq M_{\rm min}^{_\mathrm{HOD}}
    \end{cases}
\end{equation}
with $M_\mathrm{min}^{_\mathrm{HOD}}=10^{10} M_{\odot}$, the minimum halo mass to contain a galaxy. The satellite galaxy luminosity is  
\begin{equation}
    L^{\rm sat}_{\nu}(M,z)=\int_{M_\mathrm{min}^{_\mathrm{HOD}}}^{M}{\rm d}M_{\rm s}\frac{\mathrm{d}N}{{\rm d}M_{\rm s}}L^{\rm gal}_{\nu}(M_{\rm s},z)
\end{equation}
where $\mathrm{d}N/{\rm d} M_{\rm s}$ is the subhalo mass function  given in Eq.~(12) of \cite{Tinker2010}. For computational efficiency, we pre-tabulate this integral on a mass and redshift grid. Now, taking into account the contribution from both centrals and satellites, the total halo luminosity is 
\begin{equation}
    L_{\nu}(M,z)=L^{\rm cen}_{\nu}(M,z)+L^{\rm sat}_{\nu}(M,z) \,.
\end{equation}
Finally, the comoving (denoted with tilde) CIB monopole intensity \textit{at redshift} $z$ is obtained by integrating the halo luminosity over all halos beyond this redshift (see Appendix \ref{app:luminosity} for a detailed derivation): 
\begin{equation}
   \tilde{I}^{\rm CIB}_{\nu_z}(z)=\int_{z}^{z_\mathrm{max}}\mathrm{d}z^\prime \frac{c\tilde{j}_{\nu_z}(z^\prime)}{(1+z^\prime)H(z^\prime)} \quad\mathrm{with}\quad  \tilde{j}_{\nu_z}(z^\prime)=\int_{M_\mathrm{min}}^{M_\mathrm{max}} {\mathrm{d}M}\frac{\mathrm{d}N}{ \mathrm{d}M}\frac{L_{\frac{(1+z^\prime)}{(1+z)}\nu_z}(M,z^\prime)}{4\pi}\label{eq:Icibhm} \,,
\end{equation}
\noindent where we use $z_\mathrm{max}=12$ (i.e., the highest redshift where the CIB is sourced in our model) in the redshift integral and $H(z)$ is the Hubble parameter. We have checked that contributions to the CIB monopole at higher $z$ are negligible. The frequency $\nu_z$ is the photon frequency as seen by an observer at redshift $z$: a photon of frequency $\nu_z$ at redshift $z$ would appear at frequency $\nu_z/(1+z)$ today, and was at frequency $(1+z^\prime)\nu_z/(1+z)$ at $z^\prime$ (which is the frequency that enters in the evaluation of the galaxy SED in Eq.~\ref{eq:SED}). For the halo abundance, $\mathrm{d}N/\mathrm{d}M$, we use the \cite{T2010} halo mass function (i.e., a normalized version of the \cite{T08} formula)\footnote{We use a redshift-dependent normalization, i.e.,  the normalization factor $\alpha(z)$ is chosen such that $\int_{\nu_\mathrm{min}}^{\nu_\mathrm{max}}\mathrm{d}\nu f(\nu,z)b(\nu,z)=1$ with $\nu_\mathrm{min}\ll1$ and $\nu_\mathrm{max}\gg1$, where $f(\nu,z)$ is the halo multiplicity function. In practice, we interpolate the table originally provided in \url{https://github.com/simonsobs/hmvec/blob/master/data/alpha_consistency.txt}.} and integrate between $M_{\rm min} = 10^{10} \, M_{\odot}$ and $M_{\rm max} = 10^{16} \, M_{\odot}$. 
Note that we define the halo mass by the boundary enclosing an overdensity that is $200$ times the mean matter density, i.e., $M_{200m}$, for the halo mass function and the CIB SED and luminosity functions. 

As we saw in Sec.~\ref{ss:tm}, the Compton-$y$ parameter determines the amplitude of the inverse-Compton CIB distortion.  However, our simple model in the previous section assumed that all of the scattering takes place at $z=0$, whereas in reality the scattering takes place throughout the history of the universe, after the first groups and clusters form.  As seen in Fig.~\ref{fig:icib_didz}, the redshift kernel of the CIB and the Compton-$y$ fields has non-trivial overlap, i.e., CIB photons are being produced during the epoch in which hot electrons are virializing in the potential wells of galaxy clusters.  Therefore, in order to accurately predict the inverse-Compton CIB distortion, we also need to compute the halo model Compton-$y$ monopole and its redshift evolution in a self-consistent manner.  The Compton-$y$ parameter is the LOS integral of the electron pressure, $P_{e}$.  Within the halo model, the Compton-$y$ monopole $\langle y \rangle$ is given by (e.g. \cite{Hill2015,chiang2020}
\begin{equation}
\label{eq:hm_y}
\langle y \rangle = \int_\mathrm{z_\mathrm{min, y}}^{z_\mathrm{max, y}} \mathrm{d}z \frac{c\chi^2(z)}{H(z)} \,  \, \frac{\mathrm{d}y}{\mathrm{d}z}\quad\mathrm{with}\quad\frac{\mathrm{d}y}{\mathrm{d}z}\equiv \int_{M_\mathrm{min}}^{M_\mathrm{max}} \mathrm{d}M\frac{\mathrm{d}N}{\mathrm{d}M} y_0(M_\Delta,z) \,,
\end{equation}
where
\begin{equation}
    y_0(M_\Delta,z)\equiv\frac{\sigma_{T}}{m_{e}c^{2}}\frac{4\pi r_{\Delta}^3}{d_A(z)^{2}}\int_{x_{\rm min}}^{x_{\rm max}}\mathrm{d}x \, x^{2} \, P_{e}(x) \quad \mathrm{with}\quad x\equiv r/r_{\Delta}\quad\mathrm{and}\quad     r_\Delta = \left[3 M_\Delta/(4\pi  \Delta \rho_\mathrm{crit}(z)) \right]^{1/3}  \,.
    \label{eq:dydz}
\end{equation}
Here, $\chi(z) = (1+z) d_A(z)$ is the comoving distance with $d_A$ the angular diameter distance to redshift $z$, $\rho_{\rm crit}(z)$ is the critical density of the universe at $z$, and we use the fitting function for the electron pressure profile from \cite{Battaglia2012}, which is defined for $M_{200c}$, i.e., with respect to 200 times the critical density ($\Delta=200$). (See Appendix \ref{app:pp} for details on the pressure profile formula.) To convert between $M_{200m}$ (i.e., $\Delta=200\Omega_\mathrm{m}(z)$) used for the halo mass function and $M_{200c}$, we use the concentration-mass relation from \cite{bhatt2013}. In our implementation, the radial profile is integrated between $x_\mathrm{min}=10^{-5}$ and $x_\mathrm{max}=4$ (i.e., the pressure profile is truncated at four times the  radius $r_\mathrm{200c}$ as in \cite{wb2020}) while the mass integral is computed between the same mass limits as for the IR emission of Eq.~\ref{eq:Icibhm}. The pressure profile parameters from \cite{Battaglia2012} used in our calculations are listed in Table \ref{tab:B12parameters} in Appendix~\ref{app:pp}. For the total Compton-$y$ used in our toy model distortion and quoted throughout this paper ($y=1.58\times10^{-6}$, Sec. \ref{ss:tm}), we use redshift bounds $z_\mathrm{min, y}=0.005$ and $z_\mathrm{max, y}=6$. We have checked that Compton-$y$ contributions from sources at higher redshifts are negligible.\footnote{Note that this value includes only the Compton-$y$ signal from the ICM, neglecting the intergalactic medium and reionization contributions, each of which contributes roughly $\langle y \rangle \approx 10^{-7}$~\cite{Hill2015,Thiele2022}.}

\begin{figure}[!tbp]
\begin{center}
    \includegraphics[width=0.65\textwidth]{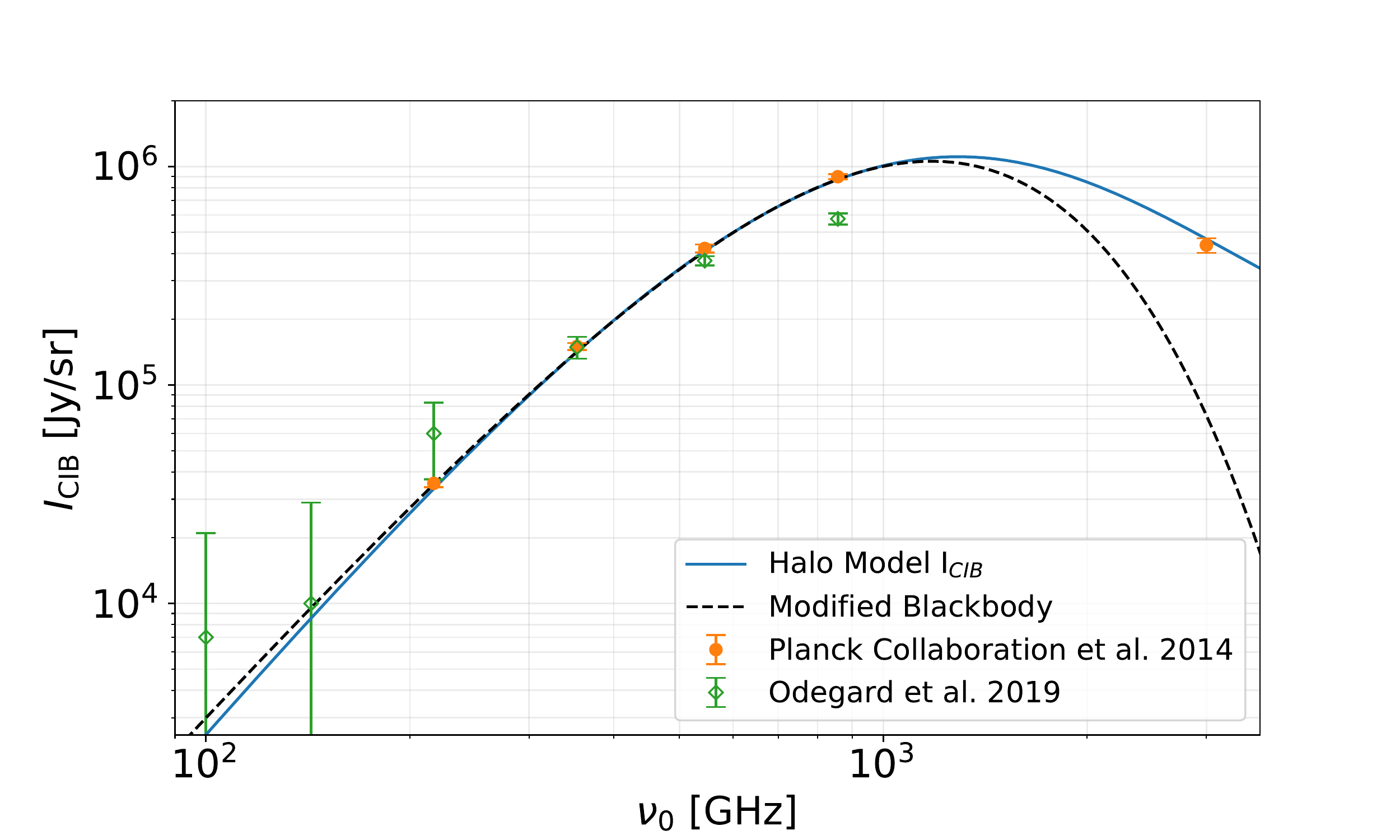}
\caption{\label{fig:matchI} MBB spectrum (dashed black) compared to the halo model-computed CIB monopole (solid blue). The MBB spectrum here is fit to the halo model CIB monopole over the frequency range 100 GHz -- 1 THz, fixing $\nu_{0}=353$ GHz and finding the best-fit values for the other MBB parameters in Eq.~\ref{eqn:modifiedN}: $T_{\rm MBB} = 12.74$ K, $\beta = 1.49$, and amplitude $A_{\rm MBB} = 6.1 \times 10^{-6} \, {\rm sr}^{-1}$. Note that these spectra do not include the inverse-Compton CIB distortion.  The CIB monopole intensity determined from \emph{Planck} HFI + IRAS ~\cite{Planck2014} (orange filled circles) and \emph{COBE}/FIRAS + \emph{Planck} HFI observations~\cite{Odegard2019} (green open diamonds) is also shown.} 
\end{center}
\end{figure}

We implement the CIB and Compton-$y$ monopole computations in the halo model code \verb|class_sz| \cite{b2018},\footnote{\url{https://github.com/asabyr/class_sz/tree/cib}, commit number 32c5f852619cf3fe488631074f2d9a5a16144603.} which performs a fast and accurate evaluation of the redshift and mass integrals using an adaptive Patterson scheme imported from  \verb|CosmoTherm| \cite{ct2011}, and is built on the underlying structure of the Boltzmann code \verb|class|~\cite{Blas2011}.\footnote{\url{https://github.com/lesgourg/class_public}, version v2.9.4 last updated on the 20th of July 2020.}  The left panel of Fig.~\ref{fig:icib_didz} shows the CIB monopole computed by integrating between various $z_{\rm min}$ and $z_{\rm max}=12$. The darkest curve ($z_{\rm min} = 0$) corresponds to the total CIB signal, which peaks at $I_{\rm CIB} \approx 10^{6}$ Jy/sr, while the faintest one shows the contribution to the CIB monopole between $z_{\rm min} = 6$ and $z_{\rm max} = 12$, which has a maximum value two orders of magnitude lower at $\approx 10^{4}$ Jy/sr.
The grey dash-dotted curves correspond to the CIB signal observed at $z=0$ (as sourced by contributions between $z_{\rm min}$ and $z_{\rm max}$), while the solid blue curves correspond to the signal seen by an observer at $z_{\rm min}$. Accordingly, the latter are shifted to higher frequencies, but the curves coincide for the calculations with $z_{\rm min}=0$.

The right panel of Fig.~\ref{fig:icib_didz} shows the CIB redshift kernels for various observational frequencies, normalized such that the total integral is unity in each case.  At the highest frequency considered (3000 GHz), most of the CIB emission originates between $0 \lesssim z \lesssim 3$, while for the lowest frequency (217 GHz), the kernel is dominated by higher redshifts ($2 \lesssim z \lesssim 6$).  We also plot the normalized Compton-$y$ redshift kernel (Eq.~\ref{eq:dydz}), which peaks near $z\approx 0$ and is dominated by contributions from $z \lesssim 2$. The evident and non-trivial overlap of the CIB and Compton-$y$ redshift kernels, which is larger at higher CIB frequencies, shows the need to account for their co-evolution in order to accurately predict the inverse-Compton scattering effect in the CIB spectrum.  We describe this calculation in the next subsection.

\subsection{Integrated CIB Spectral Distortion: Accounting for Co-Evolution}\label{ss:dicib}
In our toy model calculation in Sec. \ref{ss:tm}, we predicted the distortion of the CIB spectrum assuming a simple MBB CIB monopole undergoing inverse-Compton scattering off ICM electrons at $z=0$. However, the CIB emission and the Compton-$y$ field are sourced at overlapping redshifts, as shown in Fig.~\ref{fig:icib_didz} and described in detail in Sec.~\ref{sec:classtheory}. Therefore, in order to account for their co-evolution and calculate a more realistic signal using the halo model framework, we need to compute the differential distortion due to Compton-$y$ sources at each redshift, and then integrate over all redshifts to get the total distortion. 

\begin{figure}
    \centering
    \includegraphics[trim=110pt 0pt 110pt 0pt, width=\textwidth]{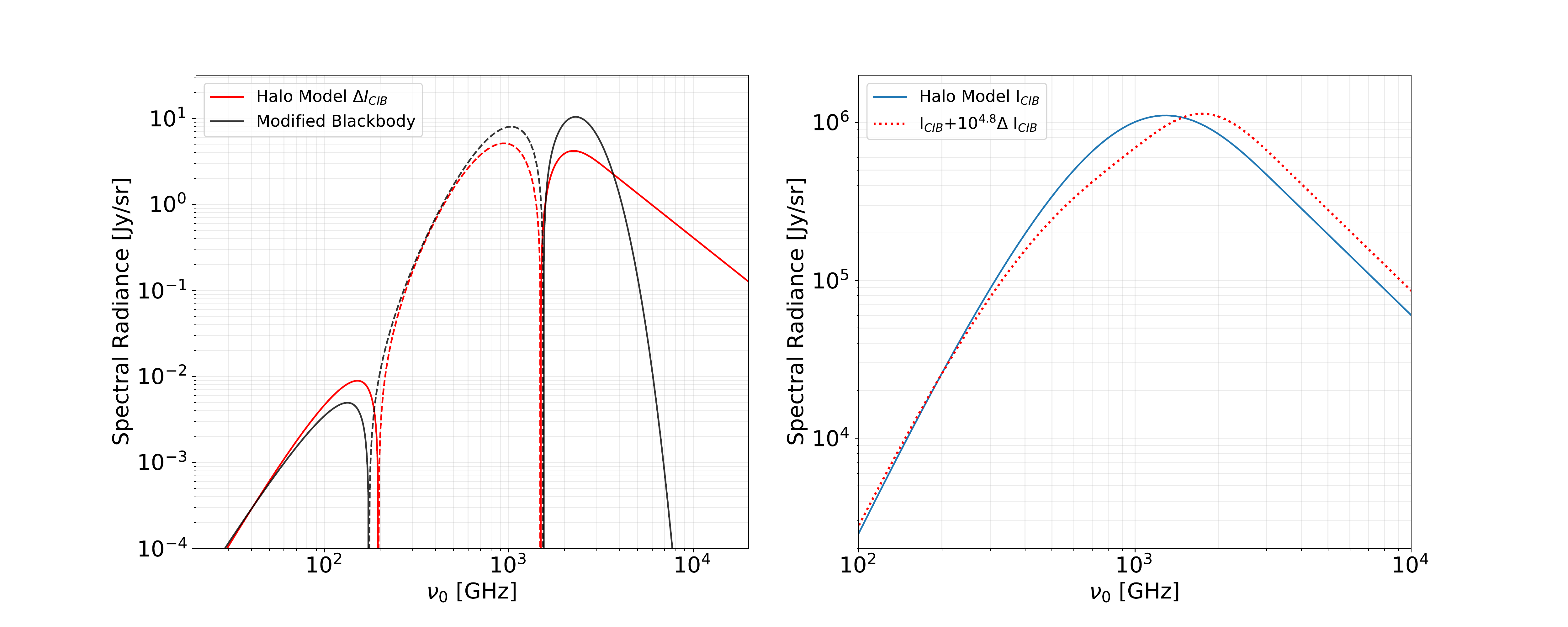}
    \caption{\textit{Left:} The inverse-Compton CIB distortion calculated using the toy model MBB SED (black) and the complete halo model for the monopole and Compton-$y$ field (red).  Negative values are plotted with dashed curves.  The distortion calculations match fairly well between 200 GHz and 2 THz, as expected from their respective CIB monopole spectra shown in Fig.~\ref{fig:matchI}, but the more accurate halo model calculation has a much larger high-frequency tail.  The maximum amplitudes are similar but lower for the halo model distortion ($\approx 4$ Jy/sr vs. $\approx 10$ Jy/sr), since it properly accounts for the fact that CIB photons generated at low redshifts are only scattered by the lower-redshift Compton-$y$ field.  The null frequencies are also slightly closer together for the halo model calculation (196 GHz and 1490 GHz) than for the MBB toy model (174 GHz and 1540 GHz). \textit{Right:} Undistorted (solid blue) and distorted (dotted red) CIB monopoles, with the distortion enhanced by a factor of $10^{4.8}$ for illustrative purposes. Here both the monopole and distortion are computed with the full halo model (rather than the MBB toy model). The two curves cross twice: at a low frequency where the CIB signal is steep ($\approx 196$ GHz) and near the peak ($\approx 1490$ GHz).}
    \label{fig:cibdistortion}
\end{figure}

To do this, we begin by expressing the Kompaneets equation (Eq. \ref{eqn:kompaneets}) in terms of specific intensity, in our case applied to the CIB monopole signal, using the relation between photon occupation number and specific intensity introduced in Eq. \ref{eqn:modifiedI}: 
\begin{equation}
\label{eq:kompaneetsI}
    \frac{\partial \tilde{I}_{\nu_{z}}^{\rm CIB}(z)}{\partial y}=\nu_{z} \frac{\partial}{\partial \nu_{z}}\left[\nu_{z}^4 \frac{\partial}{\partial\nu_{z}}\left(\tilde{I}_{\nu_{z}}^{\rm CIB}(z) \nu_{z}^{-3}\right)\right] \,
\end{equation}
where $\tilde{I}_{\nu_{z}}^{\mathrm{CIB}}(z)$ is the comoving CIB monopole intensity observed at redshift $z$ as defined earlier in Eq.~\ref{eq:Icibhm} and $\nu_{z}$ is the frequency of the photons at that redshift. This version of the Kompaneets equation allows us to calculate the distortion directly using the numerically implemented CIB model. Note that by taking the same single-scattering approximation ($y\ll1$) as in Sec.~\ref{ss:tm}, we are able to assume that the CIB monopole at redshift $z$ is the total undistorted CIB emission produced at all redshifts higher than $z$. Then the differential distortion at $z$ as seen at that redshift is given by

\begin{equation}
    \frac{\mathrm{d}(\Delta \tilde{I}^{\rm CIB}_{\nu_z}(z))}{\mathrm{d}z}=\frac{\mathrm{d}y}{\mathrm{d}z} \, \nu_z \frac{\partial}{\partial \nu_z}\left[\nu^4_z \frac{\partial}{\partial\nu_z}\left( \tilde{I}_{\nu_z}^{\rm CIB}(z) \nu^{-3}_z\right)\right]
    \label{eq:diffdist}
\end{equation}
\noindent where $\frac{dy}{dz}$ has been defined earlier in Eq.~\ref{eq:hm_y}. To find the total distortion as observed at $z=0$ at frequency $\nu_0$, we then integrate each differential distortion evaluated at $\nu_z=\nu_0(1+z)$ over redshift:

\begin{equation}
    \Delta I^{\rm CIB}_{\nu_0}(z=0) = \int_{z_{\rm i}}^{z_{\rm f}} \mathrm{d}z \frac{c\chi^2(z)}{H(z)} \frac{\mathrm{d}(\Delta \tilde{I}^{\rm CIB}_{\nu_z})}{\mathrm{d}z} \,.
    \label{eq:total}
\end{equation}
Note that we convert our final comoving distortion into proper units using $\Delta I^{\rm CIB}_{\nu_0}=\Delta \tilde{I}^{\rm CIB}_{\nu_0}/ a_{0}^{3}$, where $a_{0}=1/(1+z)=1$. In practice, the multiplicative redshift factors entering the frequency terms effectively cancel out in Eq.~\ref{eq:kompaneetsI}, so it does not make a difference whether differentiation due to inverse-Compton scattering occurs before or after the effect of CIB redshifting is included. In other words, the total distortion can be calculated following the previous steps or directly using Eq.~\ref{eq:total} and the CIB spectrum computed between $z$ and $z_{\rm max}=12$, as seen at $z=0$. For our total halo model CIB distortion, shown in Fig.~\ref{fig:cibdistortion}, we numerically integrate Eq.~\ref{eq:total} from $z_{\rm i}=0.005$ up to $z_{\rm f} = 6$, since the Compton-$y$ contribution is small at very high redshifts, as seen in Fig.~\ref{fig:icib_didz}. We sample each differential CIB monopole contribution at 100 logarithmically spaced frequencies between 5 GHz and 50 THz and use the cubic spline interpolating function from Python's SciPy library \cite{2020SciPy} to calculate the differential distortion at 1000 logarithmically spaced frequencies in that range. We verify that our results are converged with these choices, and that our discretization of the integral (with $\Delta z=0.005$) is also converged.

\section{Results}\label{sec:results}

\subsection{Comparison Between Toy Model MBB and Full Halo Model Distortion Calculations} \label{ss:comp}

Fig.~\ref{fig:cibdistortion} shows the total CIB distortion due to inverse-Compton scattering, calculated using both our analytical toy model MBB SED described in Sec.~\ref{ss:tm} and using the complete halo model prescription described in Sec.~\ref{sec:classtheory} and~\ref{ss:dicib}.  
As we anticipated based on the peak amplitude of the halo model CIB spectrum and the total Compton-$y$ parameter, the maximum positive (negative) amplitudes for our toy model and full halo model distortions are $\approx 10$ Jy/sr ($\approx -8$ Jy/sr) and $\approx 4$ Jy/sr ($\approx -5$ Jy/sr), respectively.  The positive (negative) peak amplitude in the halo model calculation is located at 2260 GHz (940 GHz).  As expected, our halo model distortion has lower amplitude than that computed using the MBB toy model, since our integration accounts for the fact that less CIB emission undergoes scattering at higher redshift, i.e., CIB photons generated at low redshifts are only scattered by Compton-$y$ sources at even lower redshifts.

\begin{figure}
    \centering
    \includegraphics[width=0.65\textwidth]{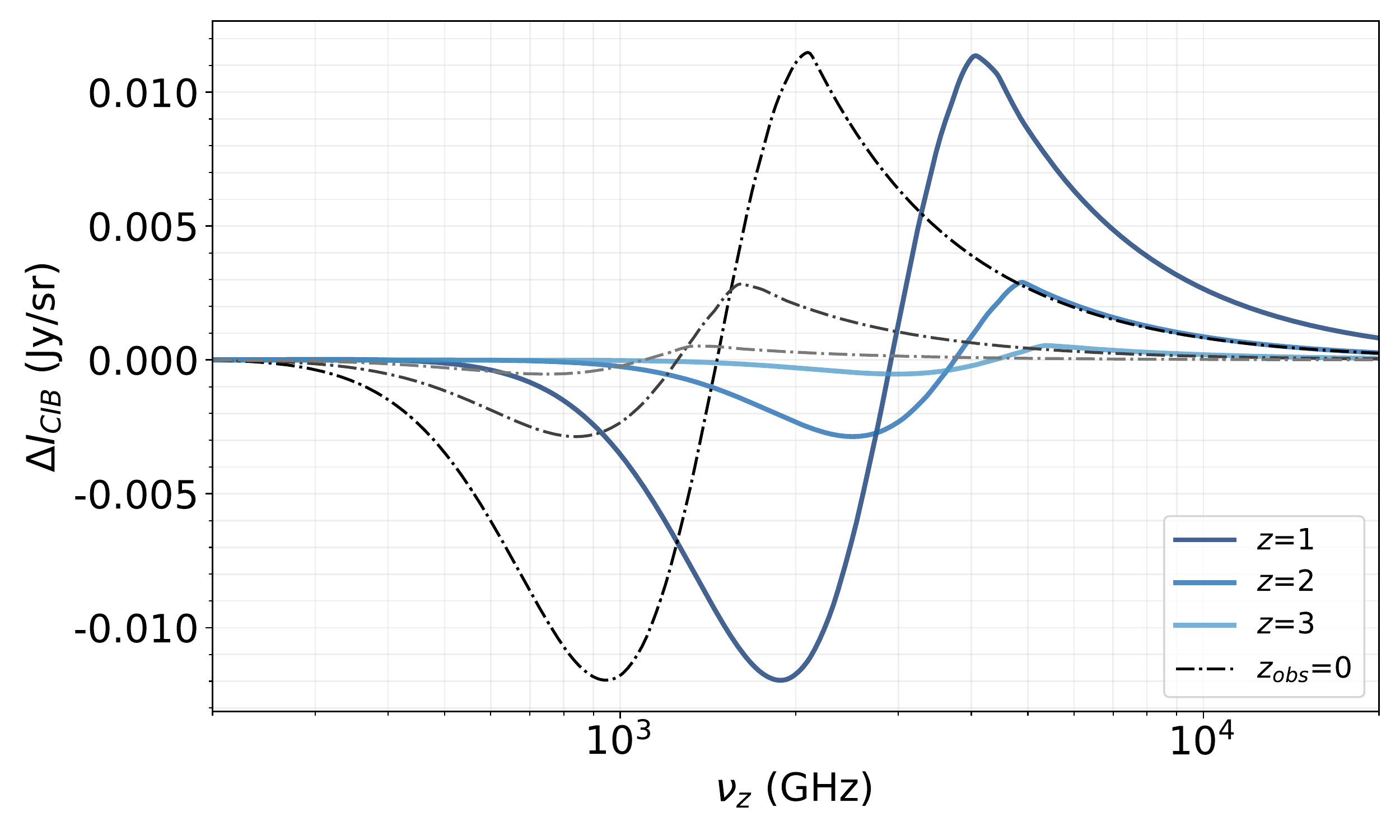}
    \caption{Differential halo model CIB distortion at several example redshifts. The blue solid curves show the distortion as seen by an observer at $z$, while the black dot-dashed curves show that seen at $z_{\rm obs}=0$.  As expected from the Compton-$y$ redshift kernel, the differential distortion contribution is larger at lower redshift, since more of the Compton-$y$ signal is located there.  We note here that the null frequencies shift as a function of redshift. The null frequencies of the dot-dashed curves move due to redshift evolution of the CIB SED (mainly because of the dust temperature redshift evolution  $T(z)=T_{0}(1+z)^{0.36}$).}
    \label{fig:diff_distortion}
\end{figure}

The shapes of the toy-model and full halo model distortions also differ. In particular, the null frequencies for the halo model distortion shift slightly closer together than those in the toy model (i.e., the lower null frequency increases, while the higher null frequency decreases). The analytic toy-model null frequencies are at 174 GHz and 1540 GHz, as calculated in Sec.~\ref{ss:tm}, while those in the halo model calculation are located at 196 GHz and 1490 GHz (see Fig.~\ref{fig:cibdistortion}).  Moreover, Fig.~\ref{fig:cibdistortion} shows that the distortion calculations differ the most at low ($\lesssim 200$ GHz) and high ($\gtrsim 2$ THz) frequencies where the MBB SED no longer describes the CIB monopole well, as seen in Fig. \ref{fig:matchI}.  Using our halo model formalism, we show the differential distortion at several example redshifts in Fig.~\ref{fig:diff_distortion}, both as seen at each redshift $z$ (blue solid curves) and at $z=0$ (dot-dashed curves). Here, the null frequencies for the observer at $z=0$ shift as a function of the dust temperature parameter in our halo model SED ($T_{d}(z)=T_{0}(1+z)^{0.36}$) and for the observer at $z$ by an additional factor of $(1+z)$. The sum of the differential distortion contributions over all redshifts yields the total distortion signal shown in red in the left panel of Fig.~\ref{fig:cibdistortion}.

\begin{figure}[!tbp]
\includegraphics[width=\textwidth]{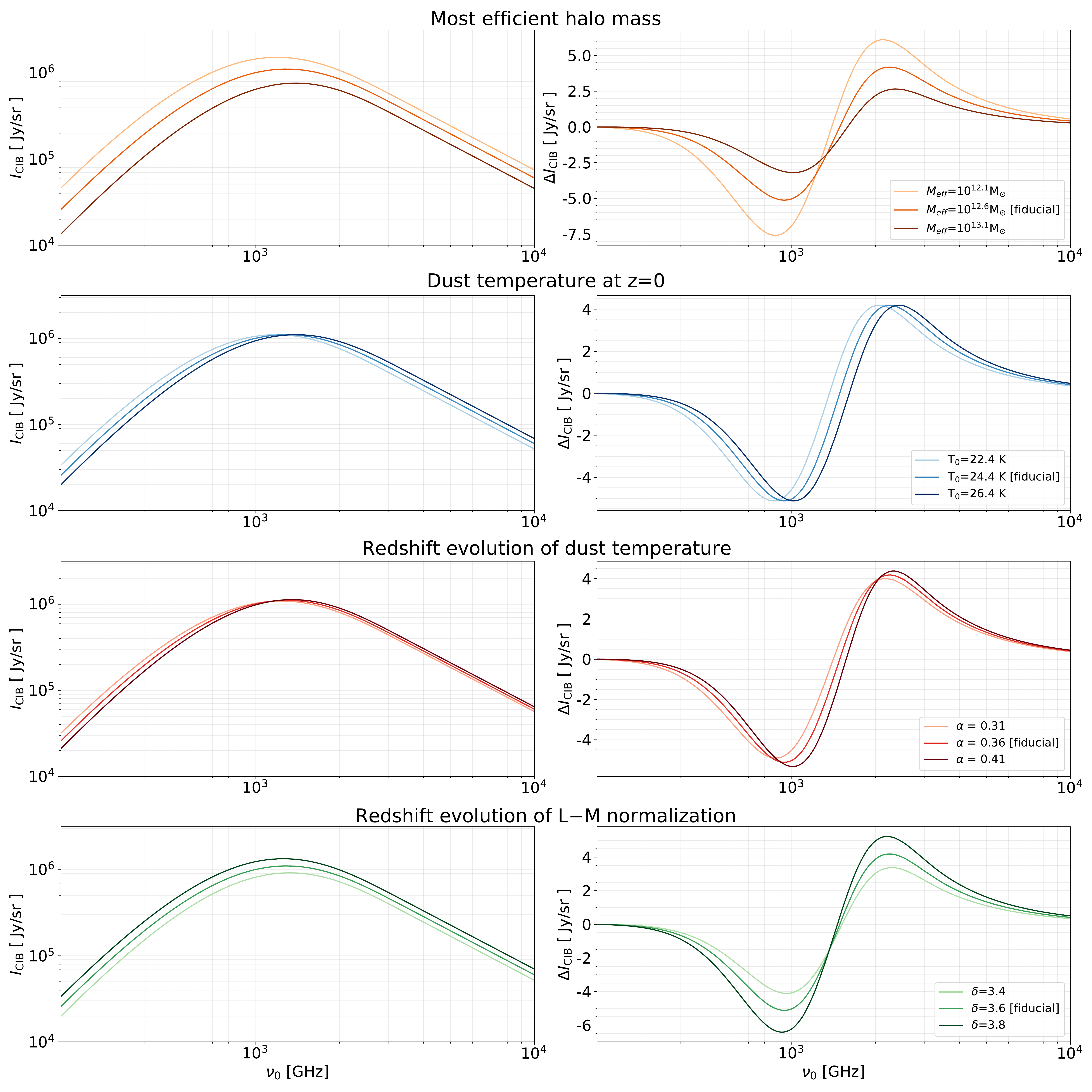}
\caption{\label{fig:varymodels} CIB monopole (left) and the corresponding inverse-Compton distortion (right) as a function of four CIB halo model parameters.  Each of the parameters (labeled in the plot legends and titles) has a different impact on the amplitude and the null frequencies of the distortion, as seen in the right panels. Their effects can largely be understood by looking at how the input CIB monopole emission changes (left panel).  See the text for further discussion.} 
\end{figure} 

We now briefly investigate some of the novel astrophysical information that can be extracted from a measurement of the inverse-Compton CIB distortion.  Both the CIB distortion computed here and the usual tSZ effect in the CMB are generated by the same Compton-$y$ sources.  However, as demonstrated in Fig.~\ref{fig:cibdistortion}, where distortions computed using the toy-model MBB and full halo model approaches are shown, the shape and the amplitude of the CIB distortion contains information about the CIB monopole and its redshift kernel.  In Fig.~\ref{fig:varymodels}, we show the effects of varying several parameters in our CIB halo model on the inverse-Compton CIB distortion.  It is interesting to note that changing each of the parameters affects the distortion differently.  For example, in the first row, when we decrease the mass of the most efficient halo mass for star formation ($M_{\rm eff}$), we effectively increase the total number of halos sourcing the CIB, so as expected the amplitude of the CIB monopole increases (more steeply at low frequencies).  The amplitude of the distortion therefore also increases and the second null frequency shifts upward due to an uneven increase of CIB emission across frequencies.  In the second row, we see that by changing the temperature of the dust at $z=0$ ($T_{0}$), we get an opposite effect on frequencies above and below the peak of the CIB emission, but the peak amplitude of the CIB monopole is unchanged.  Thus, the resulting CIB distortion has a shift in null frequency to higher (lower) frequency when the temperature is increased (decreased), but the amplitude of the distortion stays the same.  On the other hand, varying the redshift evolution of the dust temperature parameter ($\alpha$) affects both the amplitude and the null frequencies, but to a lesser extent.  Increasing the power-law index that controls the redshift evolution of the L-M normalization ($\delta$) leads to a nearly frequency-independent increase in the CIB monopole signal, and thus a corresponding coherent increase in the distortion amplitude.  The amplitude of the distortion increases, but the second null frequency does not shift.

In Fig.~\ref{fig:varymodels}, we thus see the effect of varying some of the CIB model parameters in the particular CIB model considered here.  However, the distortion calculation can be applied to other CIB models and thus used as an additional tool in understanding the origin of the CIB.  Moreover, since the integration of the halo model CIB distortion across redshift is coupled to our implementation of the CIB monopole, it can be used to study the redshift kernel of the CIB and provide information about the redshift distribution of star-forming galaxies that power the CIB emission.  In particular, as empirical knowledge of the Compton-$y$ redshift kernel $dy/dz$ improves from dedicated tSZ cross-correlation studies~\cite{chiang2020}, a measurement of the inverse-Compton CIB distortion serves as a non-trivial validation of our understanding of the CIB redshift kernel and its overlap with the Compton-$y$ kernel.

\clearpage
\subsection{Fisher Forecast} \label{ss:fisher}
To assess the detectability of the inverse-Compton CIB distortion in upcoming all-sky monopole measurements, we perform a Fisher matrix calculation for the proposed \emph{Primordial Inflation Explorer} (\emph{PIXIE}) mission~\cite{Kogut2011} and a future ESA Voyage 2050 mission~\cite{Chluba2021}, using the set-up in \cite{Abitbol2017}.\footnote{https://github.com/mabitbol/sd$\_$foregrounds} In these forecasts, we set $\Delta I_{\nu}$ to be the sky-averaged total distortion signal defined with respect to an assumed CMB blackbody SED at $T = 2.726$ K:

\begin{equation}
\label{eq:fishermodel}
    \Delta I_{\nu}=I_{\nu}^{\rm CIB}+ \Delta I_{\nu}^{\rm CIB} + \Delta I_{\nu}^{\rm CMB} +  I_{\nu}^{\rm fg},
\end{equation}
where the first two terms on the RHS are the CIB monopole, $I_{\nu}^{\rm CIB}$, and its inverse-Compton distortion, $\Delta I_{\nu}^{\rm CIB}$.Note that we use an MBB spectrum for the CIB monopole signal from \cite{Abitbol2017} for simplicity and consistency with that work, in particular the choice of free parameters varied in the Fisher forecast.  The third term in Eq.~\ref{eq:fishermodel}, $\Delta I_{\nu}^{\rm CMB}$, is the sum of the CMB spectral distortion distortion signals, and $I_{\nu}^{\rm fg}$ is the sum of all foreground contributions.  In total we have 17 free parameters with the following definitions, fiducial values, and abbreviated labels used in tables and figures:
\begin{enumerate}
    \item \textbf{CIB:} $A_{\rm CIB}=3.46\times10^{5}$ Jy/sr, MBB amplitude; $\beta_{\rm CIB}=0.86$, MBB spectral index; $T_{\rm CIB}=18.8$ K, MBB temperature; $A_{\Delta \rm CIB}=1$, dimensionless overall amplitude of our halo model CIB distortion.
    \item \textbf{CMB spectral distortions [CMB]:}
    \begin{enumerate}
        \item \textbf{Blackbody temperature deviation:} $\Delta_{\rm T}=1.2\times10^{-4}$, fractional temperature difference.
        \item \textbf{tSZ distortion:} $y=1.58\times10^{-6}$, total Compton-$y$.
        \item \textbf{Relativistic correction to tSZ}: $k_B T_{\rm e}^{\rm SZ}=1.245$ keV, $y$-weighted electron temperature.
        \item \textbf{Primordial $\mu$-distortion:} $\mu=2\times10^{-8}$, chemical potential distortion amplitude.
    \end{enumerate}
    
    \item \textbf{Foregrounds:}
    \begin{enumerate}
        \item \textbf{Galactic thermal dust [Dust]:} $A_{\rm D}=1.36\times10^{6}$ Jy/sr, MBB amplitude; $\beta_{\rm D}=1.53$, MBB spectral index; \\ $T_{\rm D}=21$ K, MBB temperature.
    \item \textbf{Galactic synchrotron [Sync]:} $A_{\rm S}=288.0$ Jy/sr, overall amplitude; $a_{\rm S}=-0.82$, power-law spectral index; $\omega_{\rm S}=0.2$, logarithmic curvature index. 
    \item \textbf{Free-Free [FF]:} $A_{\rm FF}=300$ Jy/sr, overall amplitude of the spectrum derived from \cite{Draine2011}.
    \item \textbf{Integrated CO [CO]:} $A_{\rm CO}=1$, dimensionless overall amplitude of the integrated CO template calculated using the spectra from \cite{Mashian2016}. 
    \item \textbf{Spinning dust [AME]:} $A_{\rm AME}=1$, dimensionless overall amplitude of a template signal calculated using the model in \cite{Planck2016AME}.
    \end{enumerate}

\end{enumerate}

The foreground and CMB spectral distortion models and parameters are described in detail in \cite{Abitbol2017}.  As in \cite{Abitbol2017}, we set 10$\%$ priors on the synchrotron amplitude and spectral index.  We use the same {\it PIXIE} instrument configuration with 86.4 months of spectral distortion integration time assuming $70\%$ of the sky is used in the analysis.  We set the lowest frequency bin edge at 7.5 GHz and the highest at 6 THz (note that \cite{Abitbol2017} used a highest bin cut-off at 3 THz; in our case it is advantageous to push to higher frequencies due to the non-negligible CIB distortion signal there).  Each of the frequency bins is 15 GHz wide, which gives us a total of 400 frequency channels. The parameter covariance matrix is the inverse of the Fisher information matrix, which is calculated as
\begin{equation}
    F_{ij}= \sum_{a,b}\frac{\partial (\Delta I_{\nu})_{a}}{\partial p_{i}}C_{ab}^{-1}\frac{\partial (\Delta I_{\nu})_{b}}{\partial p_{j}} \,,
\end{equation}
\noindent where indices $a,b$ denote frequency bins, $C_{ab}$ is a diagonal \emph{PIXIE} noise covariance matrix (using the noise model from \cite{Kogut2011}) and $p_{i}$ and $p_j$ are distortion and foreground parameters indexed by $i,j$ that we let vary (listed above).

\begin{figure}[!tbp]
\includegraphics[width=\textwidth]{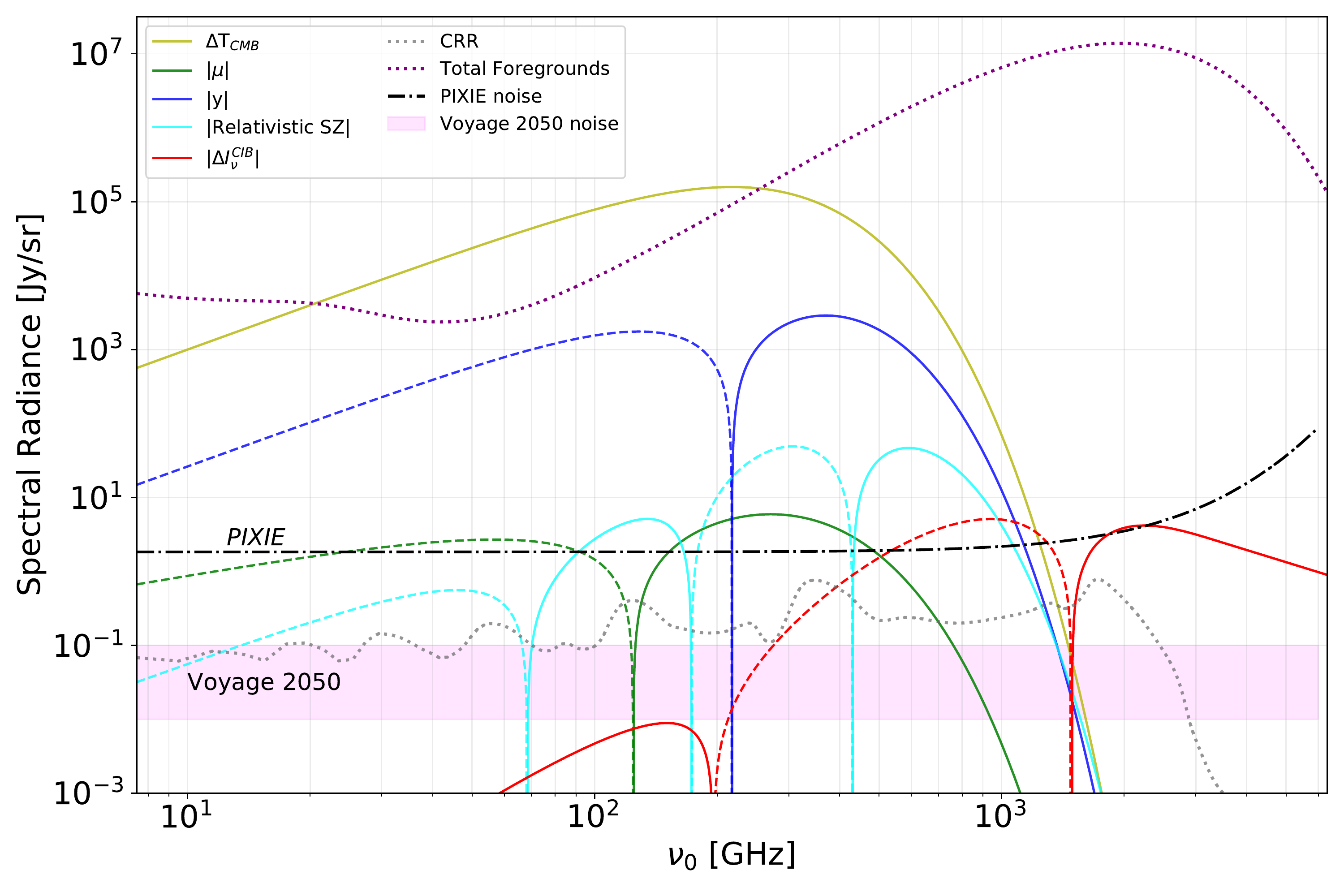}
\caption{Inverse-Compton CIB spectral distortion signal (red curve) compared to CMB spectral distortions ($\Delta T_{\rm CMB}$, Compton-$y$, $\mu$, and relativistic SZ), cosmological recombination radiation (CRR) and total foreground emission (dotted magenta).  Here, we have used $y=1.58\times10^{-6}$ (our fiducial halo model value) and $\mu=2\times10^{-8}$ for the CMB distortions.  Also shown are the noise levels for both an extended \emph{PIXIE} mission with 86.4 months of integration time and a Voyage 2050 mission with lower and upper noise limits set at 0.01 and 0.05 times the \emph{PIXIE} noise, respectively.  For the spectral distortions, negative (positive) values are indicated by dashed (solid) curves.} \label{fig:distortions} 
\end{figure}

Fig.~\ref{fig:distortions} shows a comparison of the inverse-Compton CIB distortion signal to the CMB spectral distortions, total foreground contribution, cosmological recombination radiation (CRR), and \emph{PIXIE} and ESA Voyage 2050 noise levels. 
The CIB distortion peaks roughly at $\approx 4$~Jy/sr ($\approx -5$~Jy/sr, dashed curve), while the CMB $y$-distortion peaks at $\approx 3\times 10^{3}$ Jy/sr.  The relative amplitudes follow expectations based on their relative monopole signals, which peak at $\approx 1$ MJy/sr (see Fig.~\ref{fig:cibdistortion}) and $\approx 400$ MJy/sr, respectively.  In other words, the standard CMB monopole tSZ distortion is a few hundred times larger than the CIB monopole inverse-Compton distortion.

\begin{table}[t]
\caption{\label{tab:fisher_results} Detection significance forecasts for the inverse-Compton CIB distortion using Fisher information matrix calculations. We consider different sky models and list the CIB distortion detection significance for an extended \emph{PIXIE} mission (86.4 months with 70$\%$ of the sky used for analysis) and a future Voyage 2050 spectrometer mission.  To determine the lower and upper limit forecasts for the latter, we scale the \emph{PIXIE} noise down by factors of 20 and 100, following \cite{Chluba2021}.  Here, we also assume $10\%$ priors on the synchrotron emission amplitude and spectral index as in \cite{Abitbol2017}, but excluding these priors does not have a significant effect on the CIB distortion forecasts.}
\begin{ruledtabular}
\begin{tabular}{lccccccc}
\small
Components in sky model & CIB & CIB,& CIB, CMB, & CIB, CMB, & CIB, CMB, & CIB, CMB,  & CIB, CMB, AME  \\
&&CMB& AME&AME, CO& AME, CO, & AME, CO, & CO, Sync, FF, \\
&&& && Sync & Sync, FF & Dust

\\
\colrule

\emph{PIXIE} &3.6$\sigma$&1.3$\sigma$&1.3$\sigma$&1.3$\sigma$& 1.2$\sigma$&1.2$\sigma$&0.05$\sigma$\\
Voyage 2050 (lower) & 73$\sigma$&26$\sigma$&26$\sigma$&26$\sigma$& 24$\sigma$&20$\sigma$&0.9$\sigma$\\
Voyage 2050 (upper)& 364$\sigma$&129$\sigma$&129$\sigma$&128$\sigma$& 121$\sigma$&95$\sigma$&4.6$\sigma$\\
\end{tabular}
\end{ruledtabular}
\end{table}

\begin{figure}[!tbp]
\includegraphics[width=\textwidth]{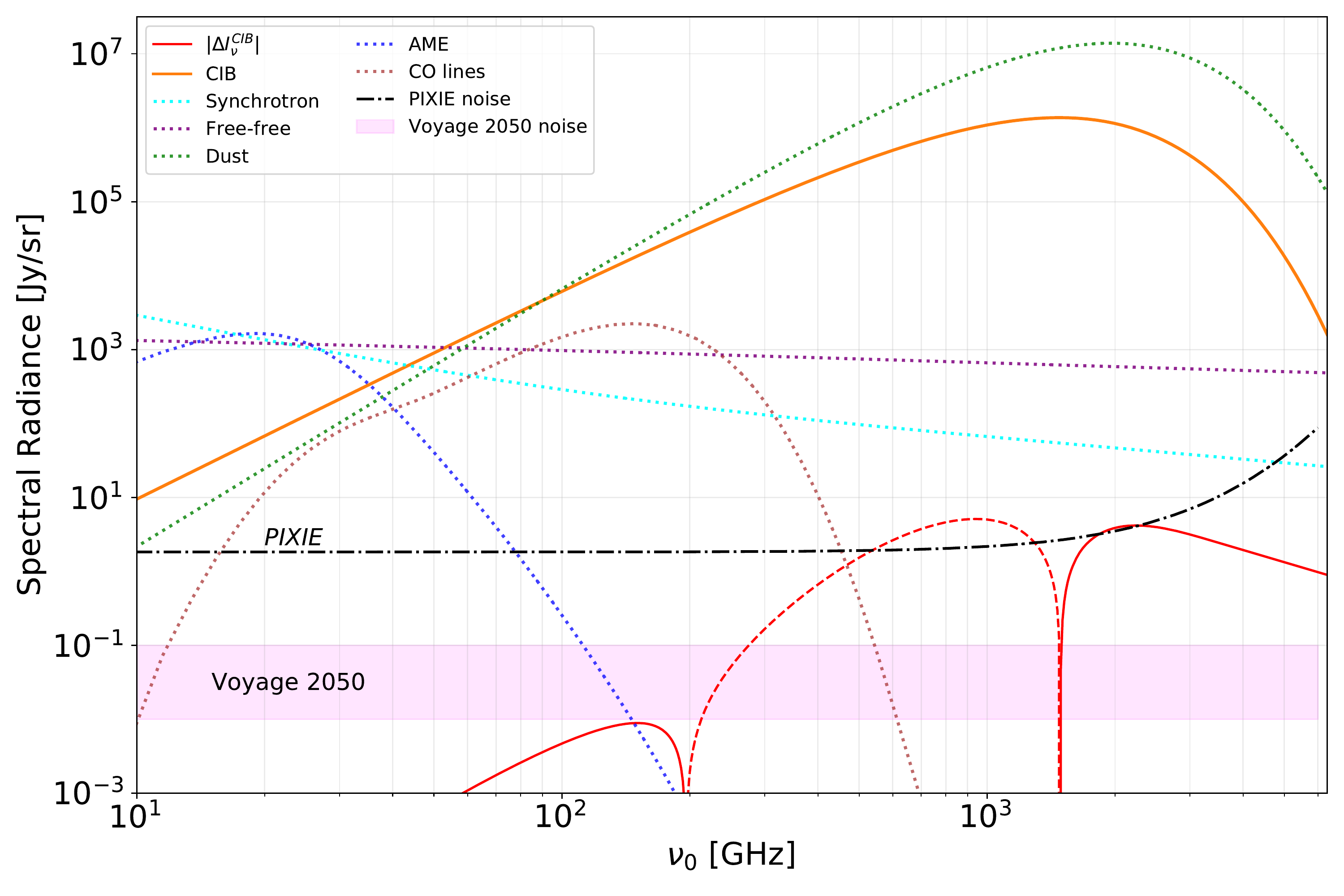}
\caption{Inverse-Compton CIB spectral distortion signal (red) compared to each of the foregrounds, and extended \emph{PIXIE} (86.4 months) and Voyage 2050 noise levels. Foregrounds are based on the models used in \cite{Abitbol2017}. The Galactic dust emission (dotted green) is the dominant foreground at the high frequencies where the CIB distortion signal is largest.}\label{fig:foregrounds}
\end{figure}

We compute the detection significance, i.e., the fiducial parameter value divided by the forecast $1\sigma$ error, of the CIB distortion for a series of sky models to determine its detectability in the presence of each of the foregrounds.  Table~\ref{tab:fisher_results} summarizes our results.  In the presence of the CIB alone, the inverse-Compton CIB distortion is measured at 3.6$\sigma$ in an extended \emph{PIXIE} mission.  However, including all foregrounds in the sky model decreases the detection significance to 0.05$\sigma$, i.e., the signal is not detected.  As expected from the frequency ranges and amplitudes shown in Fig.~\ref{fig:distortions} and Fig.~\ref{fig:foregrounds}, the CMB distortions and Galactic dust have the largest effect on the detectability of the CIB distortion. The CMB signals are important at lower frequencies, while dust dominates at high frequencies (see Fig.~\ref{fig:foregrounds}). Synchrotron emission, on the other hand, has little effect on our forecasts and therefore eliminating synchrotron priors in our analysis only decreases the forecast signal-to-noise by $\approx 12\%$ (still at $\sim 0.05 \sigma$).

Although detecting the inverse-Compton CIB distortion would be challenging for a \emph{PIXIE}-like mission due to foregrounds, we also extend our forecasts to a higher-sensitivity mission such as the ESA Voyage 2050 program \cite{Chluba2021}.  To do this, we scale the \emph{PIXIE} noise level in our calculation by a factor of 0.05 or 0.01, which correspond to the upper and lower limits of the projected Voyage 2050 noise levels \cite{Chluba2021}. Even in the presence of all foregrounds, we achieve a CIB distortion detection at $0.9\sigma$ and $4.6\sigma$ significance for Voyage 2050, as shown in Table~\ref{tab:fisher_results}.  Thus, this signal should be taken into account in forecasting for future CMB spectral distortion missions.

Finally, we note that although the inverse-Compton CIB distortion visibly overlaps at lower frequencies with the CMB spectral distortion signals, including the CIB distortion signal in the Fisher forecast does not have any significant effect on their signal-to-noise. The CIB distortion is a relatively small signal in comparison to the sum of the foregrounds at those frequencies, which have a much larger impact on the forecast signal-to-noise for the CMB distortions.

\section{Discussion \& Conclusion} \label{sec:conclusion}

In this work we have calculated the spectral distortion of the CIB monopole due to inverse-Compton scattering for the first time.  We consider both a simplified toy model in which the CIB monopole is assumed to be a modified blackbody and all of the scattering takes place at $z=0$, as well as a full halo model computation in which the co-evolution of the CIB monopole and Compton-$y$ signals are explicitly treated.  We find that the halo model CIB distortion has maximum positive and negative amplitudes of $4$ Jy/sr and $-5$ Jy/sr at 2260 and 940 GHz, respectively, and null frequencies at 196 GHz and 1490 GHz (see Fig.~\ref{fig:cibdistortion}).

Using Fisher information matrix methods, we forecast the detectability of this new inverse-Compton CIB distortion signal for an extended \emph{PIXIE} (86.4 months) mission.  We find that {\it PIXIE} has sufficient statistical sensitivity to detect this signal at $3.6\sigma$ significance in the absence of foregrounds or other sky components.  However, when realistic foreground models are included in the calculation, the forecast signal-to-noise is negligible ($0.05 \sigma$).  This is similar to the results found for the CMB $\mu$-distortion in \cite{Abitbol2017}. Improving our current knowledge of foregrounds is therefore crucial in order to observe this distortion in the near future.  Looking further ahead, scaling the noise level to ESA Voyage 2050, we find more promising results with a 0.9-4.6$\sigma$ detection significance for the CIB distortion in the presence of the CMB and all foreground components (see Table~\ref{tab:fisher_results}).

Varying some of the halo model parameters in our CIB monopole implementation has different effects on the shape and amplitude of the distortion, which suggests that the distortion can be used as an additional tool to constrain models of the CIB.  While in our work we use the CIB halo model from \cite{McCarthy2021} and \cite{Shang2012}, there are other models that would be interesting to explore.  These include the physically motivated CIB model in \cite{Maniyar2021}, which connects matter accretion onto dark matter halos to the star formation rate and thus the emissivity described in Sec.~\ref{sec:classtheory}.  Therefore it would be interesting to study this in future work as it would connect the CIB distortion more directly to the star formation history. 

Additionally, we note that our results depend on the electron pressure profile used in the model of the Compton-$y$ field, which is not currently fully constrained  (see, e.g., Fig.~1 in \cite{Thiele2022} for the range of Compton-$y$ monopole predictions from hydrodynamical simulations and \cite{chiang2020} for a range of measurements of the Compton-$y$ monopole redshift kernel). Other than the \cite{Battaglia2012} pressure profile that we used here (assuming the original fiducial parameter values reported in Table \ref{tab:B12parameters}), another common choice in the literature is the \cite{Arnaud2010} pressure profile, whose normalization depends on the hydrostatic equilibrium (HSE) mass bias. Using the latter profile, we find that the value of $\langle y \rangle$ can be up to $\sim 50$\% lower than found in our fiducial model, depending on the choice of the HSE mass bias, which remains uncertain both from theory~\cite{shi2014,Shi:2014lua,shi2015,anginelli2020} and data (e.g. \cite{b2018,Pandey:2021bdj,salvati2021combining}).
The determination of the correct model to adopt will be possible with near-future measurements of the thermal SZ effect in the CMB, such as direct measurement of the sky-averaged Compton-$y$ distortion with a spectrometer mission (e.g. \cite{maffei2021bisou}), the tSZ power spectrum (e.g. \cite{ks2002,hp2013,b2018}), or cross-correlations between tSZ and large-scale structure tracers (e.g. \cite{ka2019,chiang2020,mk2020,Pandey:2021bdj}).  In addition to this astrophysical modeling uncertainty, our calculation has some dependence on cosmological parameters, primarily $\sigma_8$ and $\Omega_m$.  However, the numerous astrophysical inputs entering our calculation of both the CIB and Compton-$y$ fields are much more uncertain than our knowledge of the cosmological parameters.

In this paper, we have focused on the inverse-Compton distortion in the CIB monopole, as a first step in investigating this new signal.  However, it is of interest to compute the distortion anisotropies in future work, as these may be observable by near-future experiments observing from the ground.  As a first step in this direction, we provide an estimate of the inverse-Compton CIB distortion power spectra in Appendix~\ref{app:as}.  We perform this calculation by taking the product of the halo model Compton-$y$ power spectrum, $C_{\ell}^{yy}$, calculated with \verb|class_sz| (\cite{b2018}, see Appendix \ref{app:as} for futher  details), and the square of our toy model MBB SED distortion as written in Eq.~\ref{eq:modified_dI} (but without the Compton-$y$ parameter there, as this has been replaced by the $y$ power spectrum).  We then estimate the signal-to-noise for detecting this power spectrum using expected sensitivity levels for the upcoming CCAT-p experiment~\cite{CCATp2021}. Fig.~\ref{fig:ccat} in Appendix~\ref{app:as} shows the noise power spectra for five CCAT-p observing frequency bands, along with our estimated signal. While the signal-to-noise that we find for the CCAT-p frequency channels is low ($\ll 1\sigma$), the CIB distortion signal peaks at higher frequencies than those observed by CCAT-p. Therefore, the forecast may be more promising at a higher frequency closer to 1 THz.  Nevertheless, higher sensitivity is clearly needed to detect the diffuse distortion signal. We note that the cross-correlation between the standard tSZ signal and the inverse-Compton CIB distortion would be easier to detect than the auto-power spectrum of the latter. However, taking into consideration the low SNR that we calculated for the CIB distortion power spectra, we most likely still need a different approach for near-term measurement of this new signal.

A more promising route to a first detection of the inverse-Compton CIB distortion signal using current technology may be to observe a known massive galaxy cluster, such as the Coma cluster at $z = 0.023$, which has a large Compton-$y$ value $y\approx 6\times10^{-4}$ \cite{PlanckComaCluster2013}, or perhaps a more distant cluster that is compact in angular extent but still possesses a large $y$ signal.  This approach is analogous to the first detections of the tSZ effect in the CMB~\cite{Birkenshaw1978a,Birkenshaw1978b,Herbig1995, Andreani1998, Komatsu1999}, which were made using large radio telescopes (e.g., OVRO) pointed at the location of massive known clusters.  As an example, for Coma we would expect the inverse-Compton CIB distortion to have a maximum positive amplitude around $\approx 2200$ Jy/sr near 2.7 THz and a maximum negative amplitude of $-2700$ Jy/sr near 1 THz.  One could also perform a stacking analysis on clusters observed at similar redshifts, for which the CIB distortion SED would be similar. The inverse-Compton CIB signal should be included in theoretical modeling for multifrequency cluster stacking measurements, such as those aiming to detect the relativistic tSZ signal. While this effect is likely negligible in analyses using Planck data (e.g. \cite{Erler2018}), it may become important in analyses using, for example, CCAT-p \cite{CCATp2021} and Simons Observatory data \cite{SO2019}.  Overall, stacking methods are likely to be more promising than a detection of the diffuse power spectrum signal considered above (similar again to the case of the standard tSZ effect, for which the power spectrum was not detected until decades after the first observations of individual objects). We leave to future work a detailed forecast of the detectability of such signals for upcoming infrared observatories.

\section{Acknowledgements}
We would like to thank Fiona McCarthy for help with the implementation of the CIB halo model and useful discussions, as well as Jens Chluba, Will Coulton, Mathew Madhavacheril, Abhishek Maniyar, Yogesh Mehta, David Spergel, and Alex van Engelen for  useful conversations. We thank Eiichiro Komatsu and the anonymous referee for useful comments on our manuscript. JCH acknowledges support from NSF grant AST-2108536.  The Flatiron Institute is supported by the Simons Foundation. This research made use of \texttt{Mathematica} \cite{Mathematica}, GSL \cite{gough2009gnu}, class\_sz \cite{b2018}, class \cite{Blas2011} and open source \texttt{Python} packages \texttt{NumPy} \cite{numpy}, \texttt{SciPy} \cite{2020SciPy} and \texttt{Matplotlib} \cite{matplotlib}. 

\appendix
\section{Flux and cosmological expansion}\label{app:luminosity}

The proper flux $S$ measured by an observer at scale factor $a_{\rm obs}$ due to a source of bolometric luminosity $L$ at scale factor $a_{\rm em}$, with comoving distance $\chi$ between emission and observation, is given by~(e.g.,~\cite{Weinberg2008,DodelsonSchmidt2021})

\begin{equation}
S=\frac{L (a_{\rm em}/a_{\rm obs})^2}{4\pi \chi^{2}a_{\rm obs}^2} \,.
\end{equation}
 
Here, one factor of $(a_{\rm em}/a_{\rm obs})$ is due to cosmological expansion, i.e., the dilution of the surface density or rate of photons as the scale factor grows with time; the second factor of $(a_{\rm em}/a_{\rm obs})$ arises from the redshifting of photons (i.e., also due to cosmological expansion), which reduces their energy.  The factor of $4\pi \chi^{2}a_{\rm obs}^2$ is the surface area of a spherical shell around the source, crossing the observer at $a_{\rm obs}$. Radiation energy density scales as $\propto a^{4}$ so we can write the following expression for the comoving flux $\tilde{S}$:
\begin{equation}
\label{eq.S}
\tilde{S}=a_{\rm obs}^4S=\frac{L a_{\rm em}^2}{4\pi\chi^2} \,.
\end{equation}
If we now define the flux density $S_\nu$ via $S = \int S_\nu \, d\nu$ and the luminosity density $L_\nu$ via $L = \int L_\nu \, d\nu$, then we can rewrite Eq.~\ref{eq.S} as
\begin{equation}
\label{eq.Snu}
\tilde{S}_{\nu}=a_{\rm obs}^3S_{\nu}=\frac{L_{\frac{(1+z_{\rm em})}{(1+z_{\rm obs})}\nu}a_{\rm em}}{4\pi\chi^2} \,,
\end{equation}
where one power of $a_{\rm em}$ and $a_{\rm obs}$ have now been absorbed into the redshifting of the photon frequency appearing in the source luminosity density. 

Specific intensity is the flux density per unit solid angle ($S_{\nu}=\int I_{\nu} \cos\theta \, d\Omega \approx \int I_{\nu} \, d\Omega$). The comoving CIB specific intensity measured at frequency $\nu_z$ by an observer at redshift $z$ is then obtained by summing the specific intensity over the contributions from all sources on the past lightcone:
\begin{equation}
\label{eq.discretesum}
\tilde{I}^{\rm CIB}_{\nu_z}(z) = \sum_i \frac{1}{4\pi}\frac{ L^i_{\frac{(1+z_i)}{(1+z)} \nu_z} a(z_i)}{4\pi \chi^2_{z,z_i}} \,,
\end{equation}
where we have defined the comoving distance between $z$ and $z_i$, $\chi_{z,z_i} = \int_z^{z_i} dz^\prime c/H(z^\prime)$, and $i$ indexes each source.  We now must take the continuous limit of this result, which involves replacing the discrete sum over sources with an integral over the comoving volume element and the halo mass function:
\begin{eqnarray}
    \tilde{I}^{\rm CIB}_{\nu_z}(z) & =& \int_z^{\infty} dz^\prime \frac{dV}{dz^\prime} \int dM \frac{dN}{dM} \frac{1}{4\pi}\frac{ L_{\frac{(1+z^\prime)}{(1+z)} \nu_z} (M,z^\prime) a(z^\prime)}{4\pi \chi^2_{z,z^\prime}} \nonumber \\
    & = & \int_\chi^{\infty} d\chi^\prime \int dM \frac{dN}{dM} \frac{1}{4\pi} \frac{ L_{\frac{(1+z^\prime)}{(1+z)} \nu_z} (M,z^\prime) \,}{(1+z^\prime)} \,,
\end{eqnarray}
where we have used the fact that $dV/dz^\prime$ here is the comoving volume element for expansion from $z^\prime$ to $z$ (not to $z=0$ as usual), and thus $dV/dz^\prime = 4\pi\chi^2_{z,z\prime} d\chi^\prime/dz^\prime$, which leads to the second line.  Also, note that $L_\nu(M,z)$ is the luminosity density of a halo of mass $M$ at redshift $z$.  We then obtain the result in Eq.~\ref{eq:Icibhm}:
\begin{equation}
\tilde{I}^{\rm CIB}_{\nu_z}(z) =a^{3}(z)I^{\rm CIB}_{\nu_z}(z)=\int_z^{\infty} dz^\prime \frac{c}{(1+z^{\prime}) \, H(z^\prime)} \int dM \frac{dN}{dM} \frac{ L_{\frac{(1+z^\prime)}{(1+z)} \nu_z} (M,z^\prime)}{4\pi} \,.
\end{equation}
The comoving specific intensity is related to proper specific intensity as $\tilde{I}^{\rm CIB}_{\nu_z}(z)=a^{3}(z)I^{\rm CIB}_{\nu_z}(z)$ as in the case of flux density. Our results are consistent with those in Ref. \cite{AcharyaChluba2022}.

\section{Pressure profile}\label{app:pp}

In Table~\ref{tab:B12parameters} we report the parameter values used in the electron pressure profile in this work.  The pressure profile is parameterized as \cite{Battaglia2012}:
\begin{equation}
 P_e(r)=C P_0 \left(\frac{x}{x_c}\right)^\gamma\left(1+\left(\frac{x}{x_c}\right)^\alpha\right)^{-\beta}\label{eq:pbat}
\end{equation}

with the dimensionful prefactor\footnote{The numerical prefactor in $C$ is obtained by converting units of $H$, $M_{200c}$, and $r_{200c}$ into $\mathrm{eV/cm^3}$, multiplying by $3\Delta/16\pi$  (with $\Delta=200$) and dividing by 1.932 to convert the thermal pressure $P_{\mathrm{th}}$ to the electron pressure $P_e$ (see Section 3 and 4 of \cite{Battaglia2012} for details).} $C = \left( 2.61051\times 10^{-18}f_b M_{200c}H^2(z)/r_{200c} \right)$ eV/cm${}^3$, where $f_b \equiv \Omega_b/\Omega_m$ is the cosmic baryon fraction, $H(z)$ is in km/s/Mpc, $M_{200c}$ is in $M_\mathrm{sun}/h$, and $r_{200c}$ in $\mathrm{Mpc}/h$. The other parameters depend on halo mass and redshift  according to $p=A_0(M_{200c}/10^{14}M_\odot)^{\alpha_m}(1+z)^{\alpha_z}$ for a generic parameter $p$, with $A_0$, $\alpha_m$ and $\alpha_z$ given in the table below.  Note that $\alpha=1$ and $\gamma=-0.3$ in Eq.~\ref{eq:pbat}, with no mass or redshift dependences~\cite{Battaglia2012}.

\begin{table}[h]
\caption{\label{tab:B12parameters} Pressure profile parameters from \cite{Battaglia2012} for their {\it AGN feedback} model and $\Delta=200$, implemented in our halo model for the Compton-$y$ field via Eq.~\ref{eq:pbat}.}
\centering
\begin{tabular}{c|c|c|c}
Parameter & $A_0$ & $\alpha_{\rm m}$ & $\alpha_{\rm z}$\\
\hline
$P_{0}$ & 18.1 & 0.154 & -0.758\\
$x_{\rm c}$ & 0.497 & -0.00865 & 0.731\\
$\beta$ & 4.35 & 0.0393 & 0.415\\
\end{tabular}
\end{table}

\section{Estimate of anisotropic distortion signal}\label{app:as}
\begin{figure}[!t]
\begin{centering}
\includegraphics[width=0.65\textwidth]{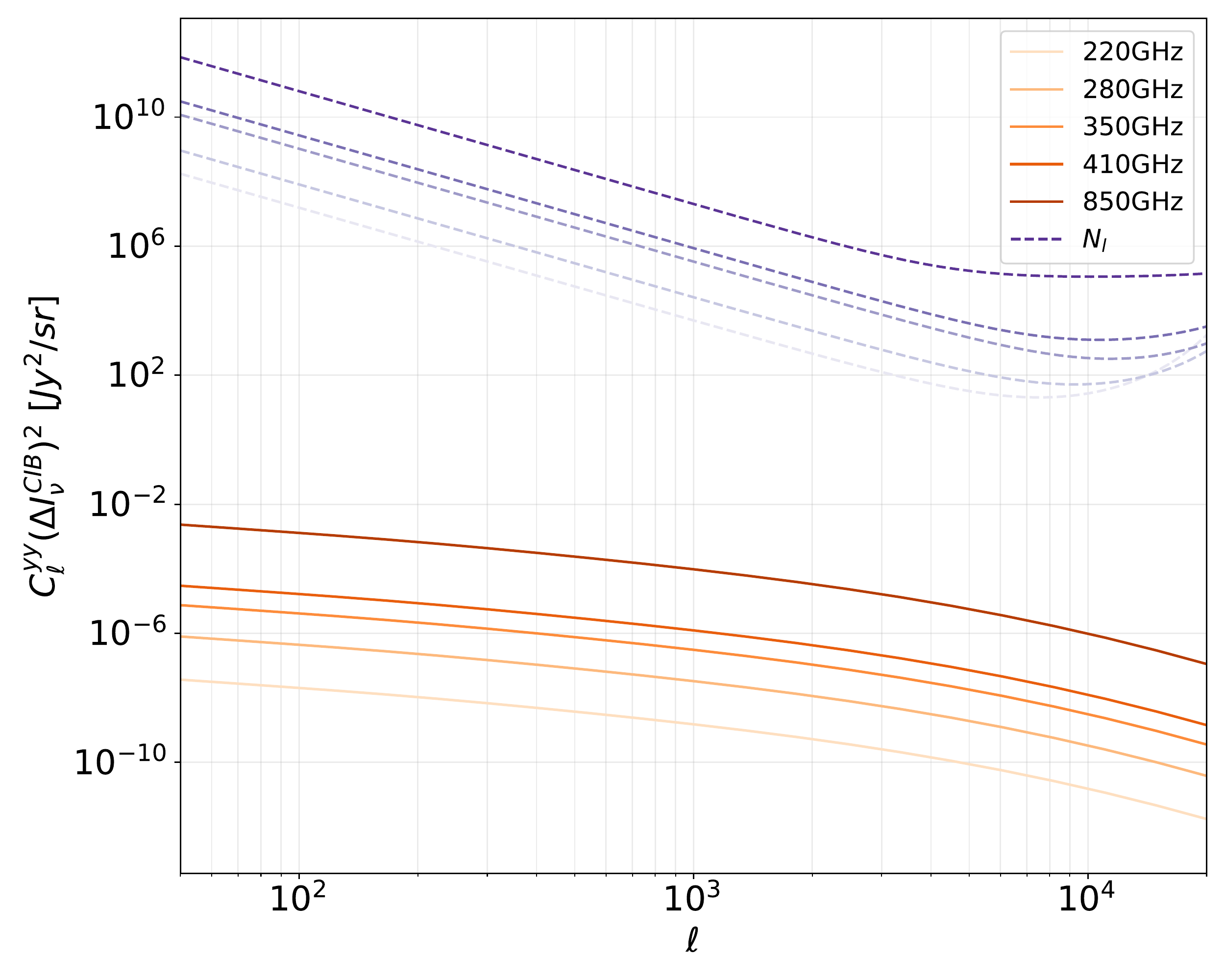}
\caption{Estimated CIB distortion power spectra (solid red/orange) compared to CCAT-p noise power spectra (dashed blue/magenta).  The inverse-Compton CIB distortion power spectra are computed using the square of our toy-model MBB SED (see Sec.~\ref{ss:tm} for details and parameter values -- note that the SED is now computed without the Compton-$y$ pre-factor in Eq. \ref{eq:modified_dI}), which is then multiplied by the Compton-$y$ power spectrum computed with standard halo model methods.  
This method is only approximate, but provides a first estimate of the order of magnitude of the signal (shown at various frequencies in red).  We compare the signal power spectra to expected CCAT-p noise power spectra (dashed blue).  While the distortion power spectra will be difficult to detect, dedicated observations of a few massive clusters in the THz band may yield a detection of the CIB distortion signal, similar to the early detections of the standard tSZ effect in the CMB at much lower frequencies.} 
\label{fig:ccat}
\end{centering}
\end{figure}

In Fig.~\ref{fig:ccat} we show a first estimate of the anisotropy power spectrum of the inverse-Compton CIB distortion, making use of the toy-model MBB distortion SED computed in Sec.~\ref{ss:tm}.  Our computation of the Compton-$y$ angular power spectrum, $C_\ell^{yy}$, is identical to that of the Compton-$y$ monopole in Eq.~\ref{eq:hm_y}, except that $y_0$ is replaced by the multipole-dependent profile $y_\ell$ given by \cite{ks2002}:
\begin{equation}
    y_\ell(M_\Delta,z)\equiv\frac{\sigma_{T}}{m_{e}c^{2}}\frac{4\pi r_{\Delta}^3}{d_A^{2}}\int_{x_{\rm min}}^{x_{\rm max}}\mathrm{d}x \, x^{2} \, \mathrm{sinc}(k_\ell ar)P_{e}(x) \quad \mathrm{with}\quad k_\ell\equiv(\ell+1/2)/\chi\quad\mathrm{and}\quad r=xr_\Delta.
\end{equation}
In \verb|class_sz|, the integral over $x$, that is the Hankel transform of the pressure profile, is evaluated using \verb|gsl|'s QAWO scheme.  We include both the one-halo and two-halo contributions to $C_{\ell}^{yy}$ (see, e.g.,~\cite{Komatsu1999,ks2002,hp2013}).  To obtain the frequency-dependent inverse-Compton CIB distortion power spectrum, we then multiply $C_\ell^{yy}$ by the square of the toy-model MBB distortion SED computed in Sec.~\ref{ss:tm}, i.e., the SED in Eq.~\ref{eq:modified_dI} with $y$ set to unity.  This approach is not correct in detail, as we are implicitly assuming that the CIB is a true primordial background field like the CMB.  A correct, complete calculation would involve a detailed treatment of the co-evolution of the CIB and Compton-$y$ fields, as done for the CIB monopole distortion via the halo model in Sec.~\ref{sec:classtheory} and~\ref{ss:dicib}.  Nevertheless, the approximate treatment here is useful to get a sense of the expected order of magnitude of the signal, which can be compared to noise levels for upcoming anisotropy measurements in the THz range (e.g., CCAT-p).  We leave a full calculation of the anisotropy signal to future work.

\bibliography{main}
\end{document}